\documentclass[11pt,a4paper]{article}
\usepackage{ska_memo} 
\usepackage{graphicx}
\usepackage{natbib}
\usepackage{xcolor}

\newcommand{\figref}[1]{Figure~\ref{#1}}

\newcommand{\bigdoctitle}{\MakeUppercase{Applications of antenna-level buffering}\xspace}

\newcommand{\docauthor}{A.\ Nelles (DESY, FAU Erlangen), J.D.\ Bray (University of Manchester), C.W.\ James (Curtin University, CAASTRO), High Energy Cosmic Particles (SKA Focus Group)\xspace} 

\newcommand{\contactauthor}{anna.nelles@desy.de}

\begin{document}


\thispagestyle{empty} 
\setlength{\extrarowheight}{1pt} 

\begin{figure*}[h]
 \centering
 \includegraphics[width=0.5\textwidth]{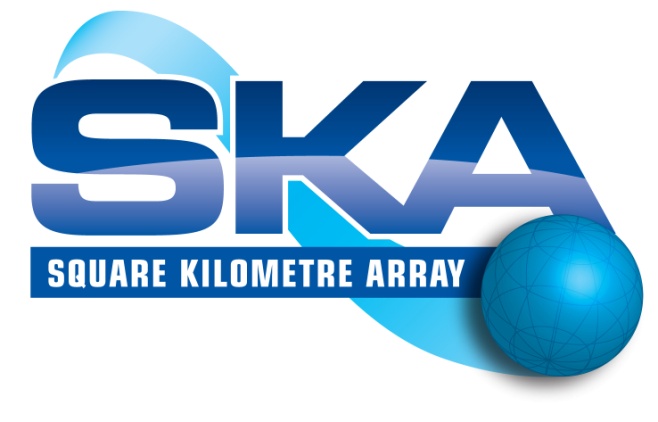}
\end{figure*}

\begin{center}
 \fontsize{22}{24}\selectfont \sffamily\bigdoctitle
\end{center}

\vspace{1cm}

\noindent\normalsize{Authors: \docauthor}\\
\normalsize{Contact: \dotfill \contactauthor}

\skasummary

The purpose of this document is to discuss applications of the antenna-level buffering capability being implemented in SKA-LOW.  In addition to their scientific motivation -- to detect and study cosmic rays interacting in the atmosphere -- these buffers provide access to low-level data for engineering and development work.  Experience has shown that antenna-level buffered data can assist with antenna calibration, localising RFI and diagnosing fundamental hardware problems.  In this document we describe several of these applications, with close reference to experience with LOFAR, ACTA, Parkes and the OVRO-LWA. 

\skatableofcontents

\skalistoffigures


\skalistofabbreviations
\begin{basedescript}{\desclabelstyle{\pushlabel}\desclabelwidth{6em}}
 \item[ATCA] Australia Telescope Compact Array \vspace{-0.2cm}
 \item[CABB] Compact Array Broadband Backend \vspace{-0.2cm}
 \item[CBF\_LOW] Central Beamformer \vspace{-0.2cm}
 \item[CRKSP] Cosmic Ray Key Science Project at LOFAR \vspace{-0.2cm}
 \item[CSP] Central Signal Processor\vspace{-0.2cm}
 \item[FRB] Fast Radio Bursts\vspace{-0.2cm}
 \item[GPS] Global Positioning System \vspace{-0.2cm}
 \item[HBA] High-Band Antenna (at LOFAR, 110-190 MHz) \vspace{-0.2cm}
 \item[HECP] High Energy Cosmic Particles \vspace{-0.2cm}
 \item[LFAA] Low-Frequency Antenna Array\vspace{-0.2cm}
 \item[LOFAR] LOw Frequency ARray \vspace{-0.2cm}
 \item[LUNASKA] Lunar Ultra-high energy Neutrino Astrophysics with the SKA  \vspace{-0.2cm}
 \item[LBA] Low-Band Antenna (at LOFAR, 10-90 MHz) \vspace{-0.2cm}
 \item[OVRO-LWA] Owens Valley Radio Observatory - Long Wavelength Array\vspace{-0.2cm}
 \item[PCB] Printed circuit-board \vspace{-0.2cm}
 \item[RFI] Radio Frequency Interference \vspace{-0.2cm}
 \item[SKA] Square Kilometre Array \vspace{-0.2cm}
 \item[SKAO] SKA Organisation \vspace{-0.2cm}
 \item[SKA-LOW] SKA LOW-frequency array \vspace{-0.2cm}
 \item[TPM] Tile processing module
\end{basedescript} 

\skalistofsymbols
\begin{basedescript}{\desclabelstyle{\pushlabel}\desclabelwidth{6em}}
 \item[$E_{\rm CR}$] Cosmic-ray energy \vspace{-0.2cm}
 \item[$T_{\rm sys}$] System temperature \vspace{-0.2cm}
\end{basedescript} 





   




\newpage

\noindent This note is structured as follows: We first introduce technical context for those not familiar with the structure of the planned SKA buffers or the science case for buffers. We then report on experiences at LOFAR, the OVRO-LWA, ATCA, and Parkes concerning RFI, system calibration and system health. This is followed by a discussion of how the use of buffers may be increased in the future. 

\section{Technical context}

\begin{figure}
 \begin{center}
  \includegraphics[width=0.8\linewidth]{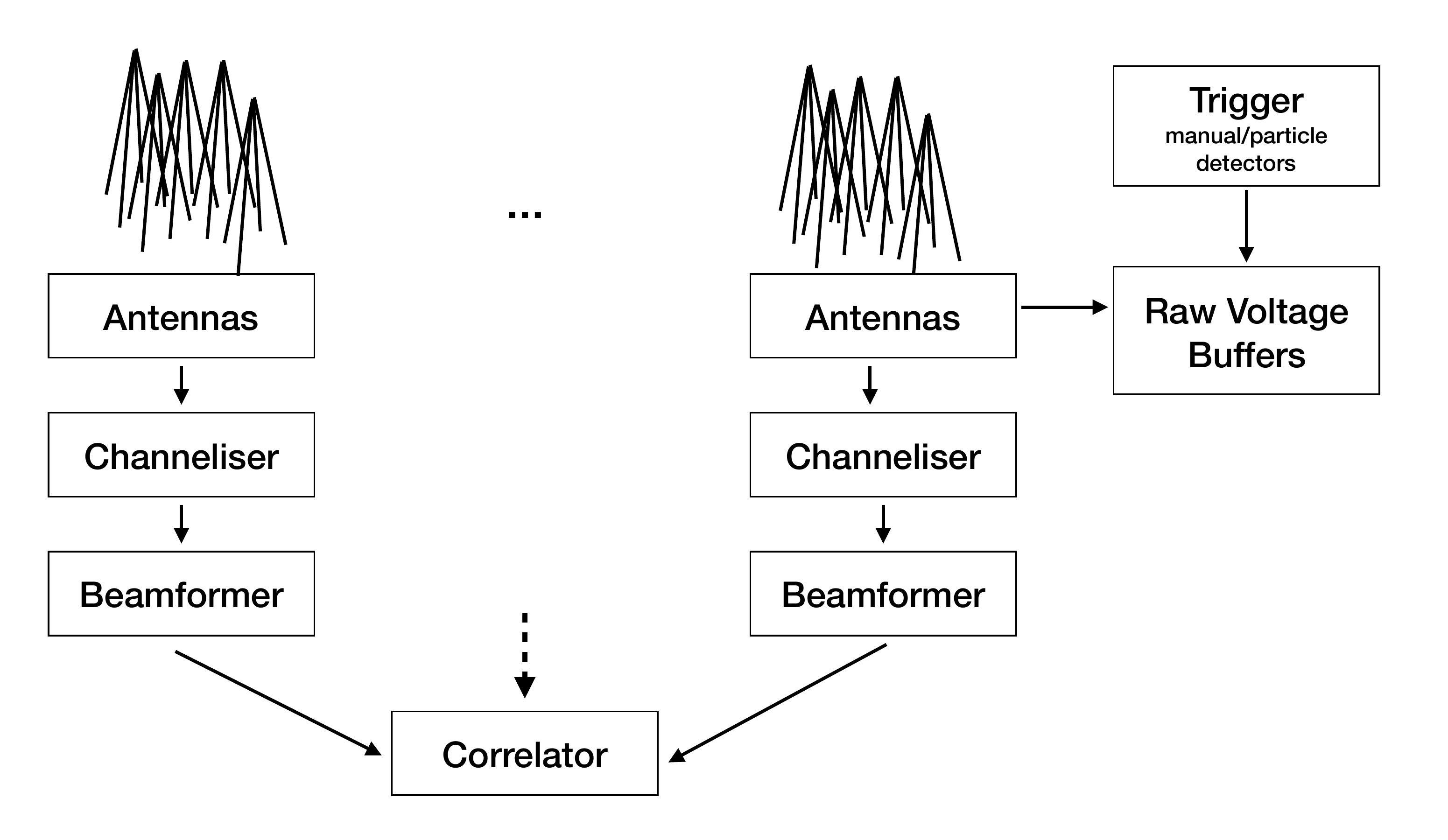}
 \end{center}
 \caption[Schematic illustrating antenna-level buffers]{A simple schematic showing antennas and beamforming taking place. It highlights the position of the buffers for the raw data in the data stream of a telescope system.} 
 \label{fig:schematic}
\end{figure}

SKA-LOW is planned to operate in a hierarchical fashion, with the outputs of the 256~antennas of each LFAA station channelised and beamformed by the Tile Processing Modules (TPMs), with the beamformed station data sent either to the Central Signal Processor (CSP) for correlation, or the Central BeamFormer (CBF\_LOW) for pulsar and single-pulse (i.e.\ FRB) searches.  This document relates to the buffering of raw voltages from each antenna (see \figref{fig:schematic}), prior to channelisation, as described in ECP~150010. The data at this level will be baseband 12-bit voltages sampled at 800\,MHz from each LFAA antenna, although only a subset of these antennas may be available for buffering at any given time.

The scientific motivation for these buffers is for the detection of sub-microsecond-scale radio pulses from high-energy cosmic rays interacting in the atmosphere above the telescope \citep[e.g.][]{buitink2016,2015aska.confE.148H}.  While such buffering capability could in principle be used for retrospective detection of transient astronomical events such as fast radio bursts outside the station-level field of view, in practice the memory requirements prohibit a buffer length longer than at most a few seconds, versus the minutes required for this latter application. 
One should also note that recently buffers have started being used for lightning research at LOFAR, where unprecedented accuracy has been reached \citep{Hare2018}.

Beyond its scientific application, antenna level buffers have a range of possible applications for identifying problems and faults with a telescope.  This document will review these applications, with reference to experience using similar buffers on other telescopes, to inform the implementation of this buffering capability with SKA-LOW. In particular, we draw on experience from the LOFAR Cosmic Rays Key Science Project; the Lunar UHE Neutrino Astrophysics with the Square Kilometre Array (LUNASKA) collaboration; and the Owens Valley Long Wavelength Array (OVRO-LWA). 

Throughout, keep in mind that the results presented are the result of work designed to detect cosmic ray signals, and as such, investigations typically ceased once the precision required to identify these signals unambiguously was achieved. We expect far more precise results to be attainable from dedicated investigations. 

\section{Characteristics of cosmic ray signals}

One of the main scientific purposes for the buffers is the detection of cosmic rays, as was most successfully shown with LOFAR \citep{schellart2013,Nelles2014,buitink2016}. The radio signals of air showers, induced by cosmic rays of roughly more than $10^{16}$ eV, manifest themselves as short, non-repeating, relatively strong nanosecond-scale pulses. The signals are broad-band with a characteristic frequency spectrum, with the maximum frequency depending to first order on the location of the observer with respect to the air shower axis. The signals are polarized, where the polarization angle depends on the arrival direction of the cosmic ray. As cosmic rays arrive (almost) isotropically on Earth, all directions (and thereby signal polarizations) are sampled over time. The electromagnetic field amplitudes of the pulses scale linearly with the energy of the primary cosmic ray. At the lower detection threshold of about $10^{16}$ eV the pulse power is roughly equal to the power in the diffuse Galactic synchrotron emission at 30-80 MHz. Cosmic rays reach energies up to $10^{20}$ eV, which cause pulses of an amplitude of a factor of $10^4$ above the noise floor. However, as the flux of cosmic rays falls steeply ($\sim E_{cr}^{-3}$) with energy, smaller signals are significantly more likely. At an area of 1 km$^2$ roughly one cosmic ray with energy above $10^{16}$ eV arrives each minute. 

\begin{figure}
 \begin{center}
  \includegraphics[width=0.49\linewidth]{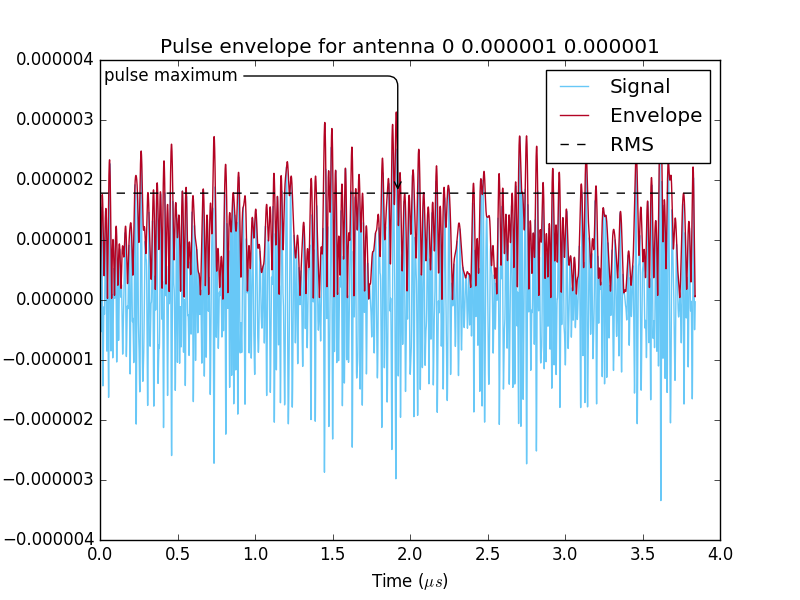}
  \includegraphics[width=0.49\linewidth]{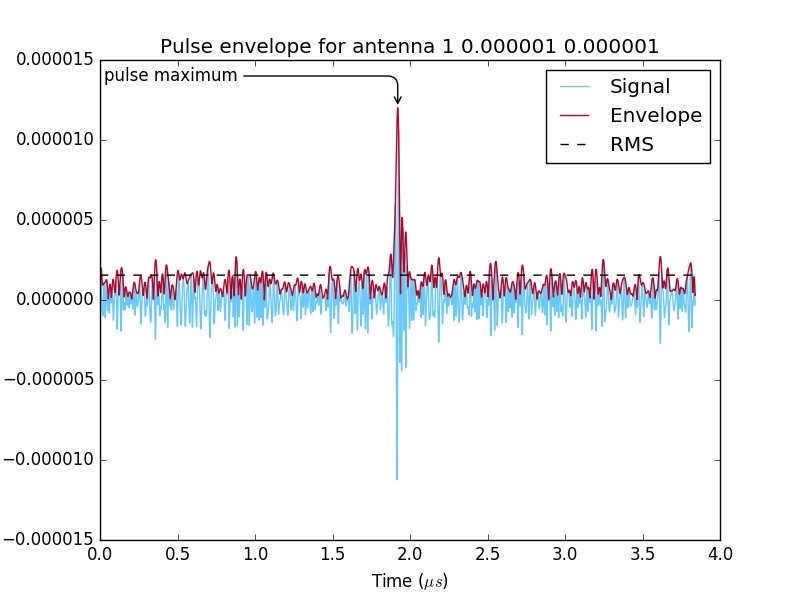}
 \end{center}
 \caption[Example of cosmic ray pulse]{Detected cosmic ray signal in two perpendicular elements (of the same dipole antenna) at LOFAR. The left figure illustrates one of the weaker pulses (antenna 0), while the right side shows a stronger, yet typical example (antenna 1). Shown are both the original raw signal from the buffers (amplitude in system units) and the envelope. Due to the strong polarization of the pulse it is only easily visible in one element (antenna 1). The images are standard output files of the cosmic ray pipeline \citep{schellart2013}.}
 \label{fig:CR_0}
\end{figure}

\begin{figure}
 \begin{center}
  \includegraphics[width=0.49\linewidth]{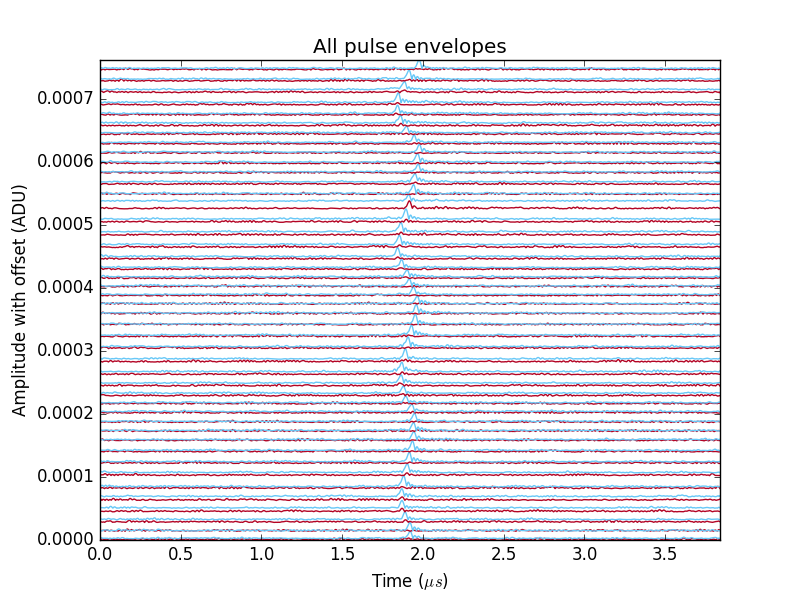}
  \includegraphics[width=0.49\linewidth]{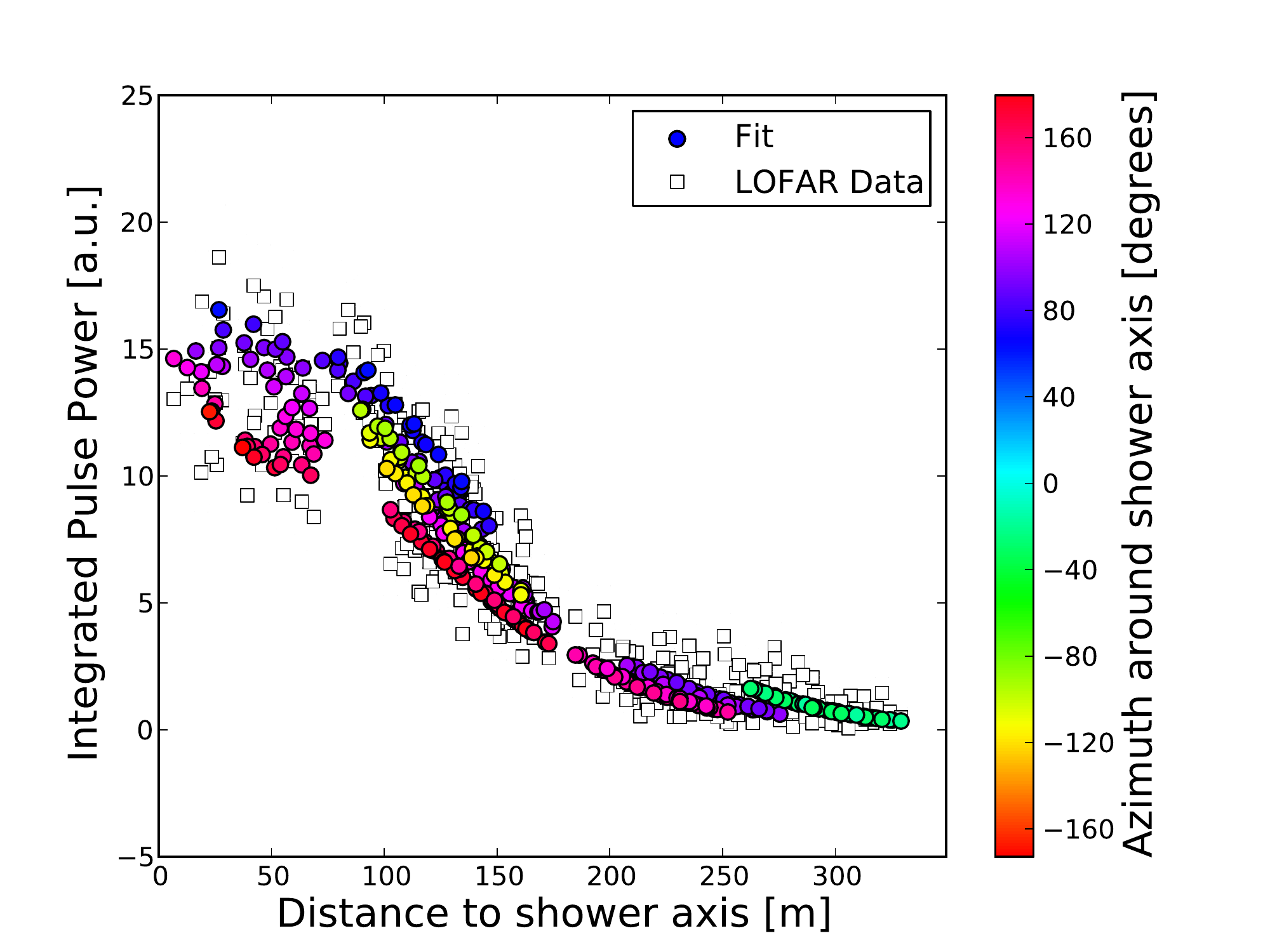}
 \end{center}
 \caption[Distribution of cosmic ray pulse amplitudes]{Left: Pulse envelopes of the raw buffer data as recorded in all dipole antennas (LBA Outer) of one LOFAR station. The arrival direction of the cosmic ray leads to a time delay for each signal. The LOFAR dipoles are numbered from inside to outside of a station in a circular fashion, which results in the visible oscillating pattern as function of antenna number (lowest line corresponds to antenna 0). Right: Pulse power as a function of distance to the shower axis and azimuth angle around the shower axis (colour). This image illustrates the strong variation of power seen in one recorded air shower. The precise shape is a result of the interference of different emission mechanisms and depends on the bandpass in which the signal is observed. Simulations predict this pattern to a high accuracy (e.g.~\cite{buitink2016}).}
 \label{fig:CR_1}
\end{figure}

Examples of the air shower signal characteristics as detected with LOFAR are given in Figures \ref{fig:CR_0} and \ref{fig:CR_1}. More details about the reconstruction procedure and event identification can be found in \citep{schellart2013}.

\section{Dealing with RFI}

The high-time-resolution, unprocessed data available through buffers provides a unique probe of radio-frequency interference (RFI). Typically, the RFI situation is unique to each experiment/location. Here, we report the experiences of cosmic ray experiments at LOFAR, ATCA, and OVRO-LWA. In the case of LOFAR, we group this capability into two classes: narrow-band RFI, and broadband (impulsive) RFI.

In the case of broadband RFI, the high-time-resolution data provided by each antenna element translates into an excellent directional/positional resolution, allowing the exact locations of RFI to be identified. It should be noted that this RF localization is only available if there is the technical option to trigger on the incoming data itself (i.e.\ threshold crossing) instead of following an external trigger. Depending on the rate of RF, it will be otherwise rather unlikely to collect enough RF pulses coincidentally.

\subsection{LOFAR -- Narrowband RFI}

The LOFAR Transient Buffer Boards record dual-polarisation data (16 bits at 200 MHz) from up to 48 LBAs simultaneously. Upon receiving an external trigger (e.g.\ a cosmic ray signature from the particle detector array \citep{Thoudam2014}, or a manual trigger for diagnostic purposes) 2.1~ms are returned, although in principle the full length of 5~s can be returned. This allows for an arbitrary time--frequency transform to be applied in order to identify narrowband RFI.

\begin{figure}
 \begin{center}
\includegraphics[width=0.49\linewidth]{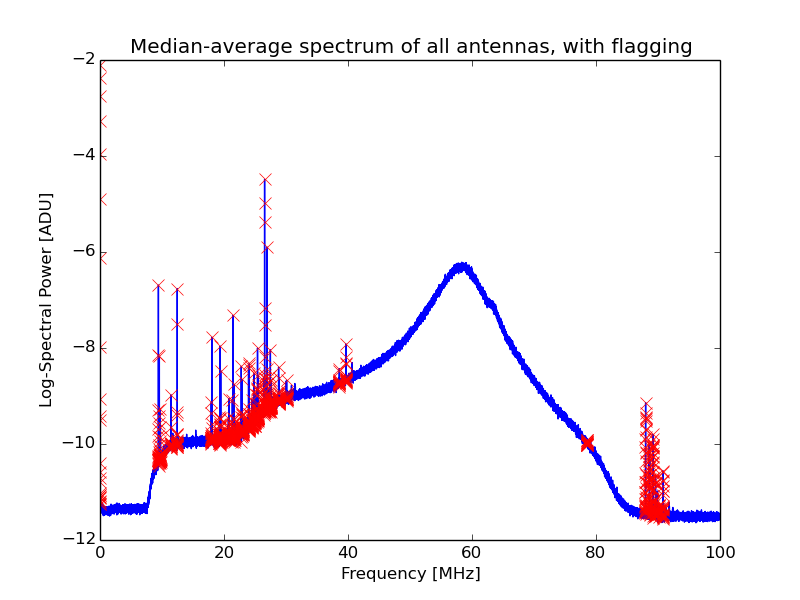}
  \includegraphics[width=0.49\linewidth]{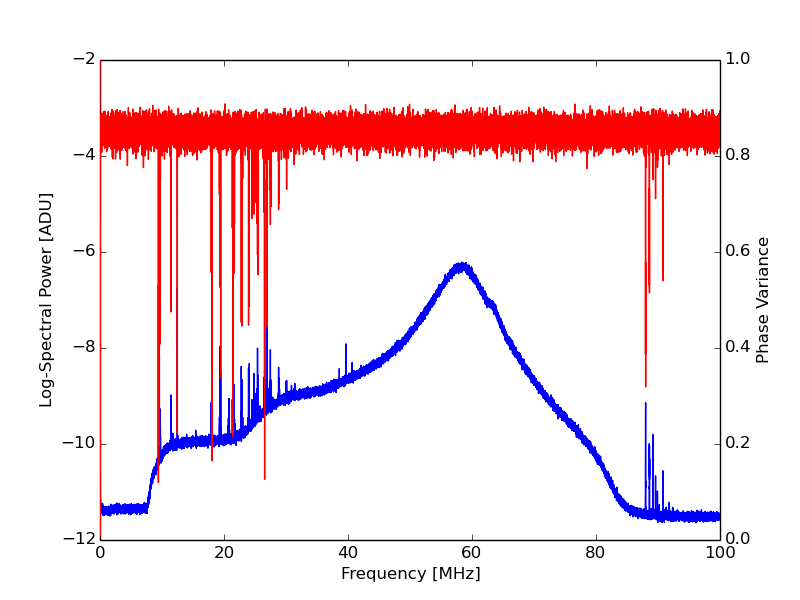}
\includegraphics[width=0.49\linewidth]{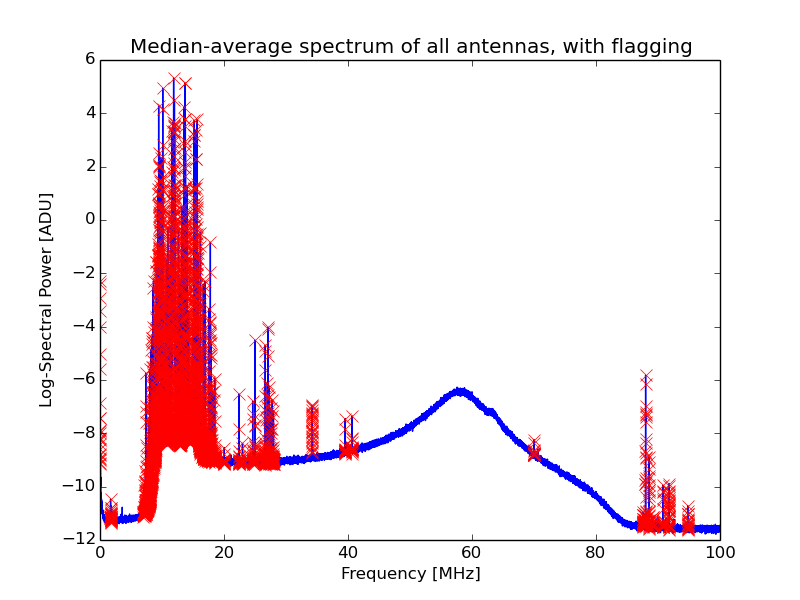}
\includegraphics[width=0.49\linewidth]{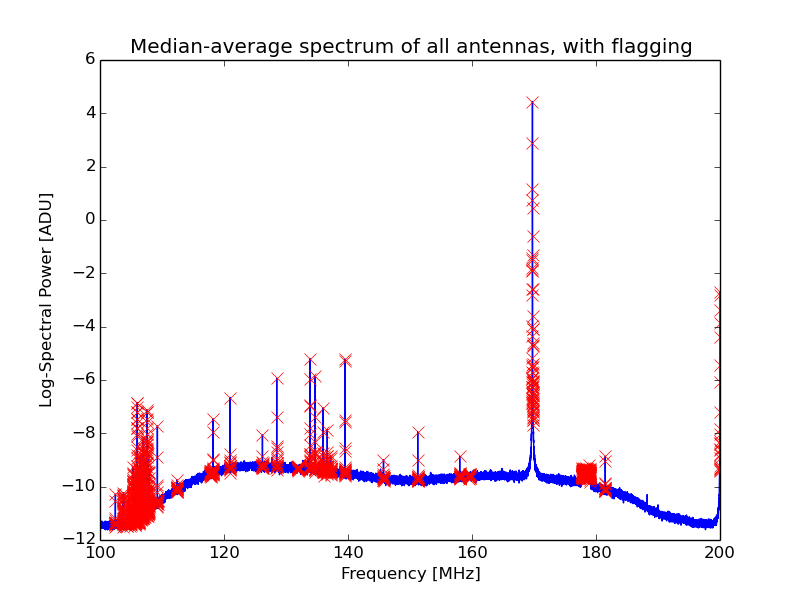}
 \end{center}
 \caption[Example of RFI spectrum at LOFAR and cleaning mechanism]{Top row: Relatively clean LOFAR LBA spectrum (blue line) with flagged RF channels (red crosses). The phase variance of each frequency bin is used to find a bandpass-independent measure for RF lines as shown on the top right (details in \citep{Corstanje2016}).
 Bottom row: One example of a less clean LBA spectrum and a typical HBA spectrum, both with flagging from the above mentioned algorithm. All images are standard results produced by the cosmic ray pipeline for every recorded cosmic ray trigger.}
  \label{fig:lofar_rfi_spectrum}
 \end{figure}

As detecting cosmic rays relies on identifying the pulse above the background noise in the full available bandwidth, the efficiency is extremely sensitive to narrowband RF noise at any location within the passband. At LOFAR, a very versatile algorithm has been developed to flag every RF line within the cosmic ray data. As shown in \citep{Corstanje2016} it detects any narrowband lines with high efficiency for every triggered data-dump. From a practical perspective, due to this approach a standard product of the cosmic ray pipeline is a continuous monitoring of the RF situation, which could allow for an early warning system for new RF lines or a database of transmitters that need to be pre-cleaned. Due to the limited size of the data, it is possible to make a dynamic spectrum of the entire life-time of LOFAR in less than an hour. Three examples of the RF flagging for the LOFAR LBAs and HBAs are shown in \figref{fig:lofar_rfi_spectrum}. 

\subsection{LOFAR -- Broadband RFI}

Next to narrowband RF, broadband RF is an issue for all radio-telescopes. In principle, cosmic rays themselves are an RF background for astronomical observations, however, their relative rarity makes their effect negligible. The situation is, however, different for terrestrial RF sources.

\begin{figure}
 \begin{center}
  \includegraphics[width=0.7\linewidth]{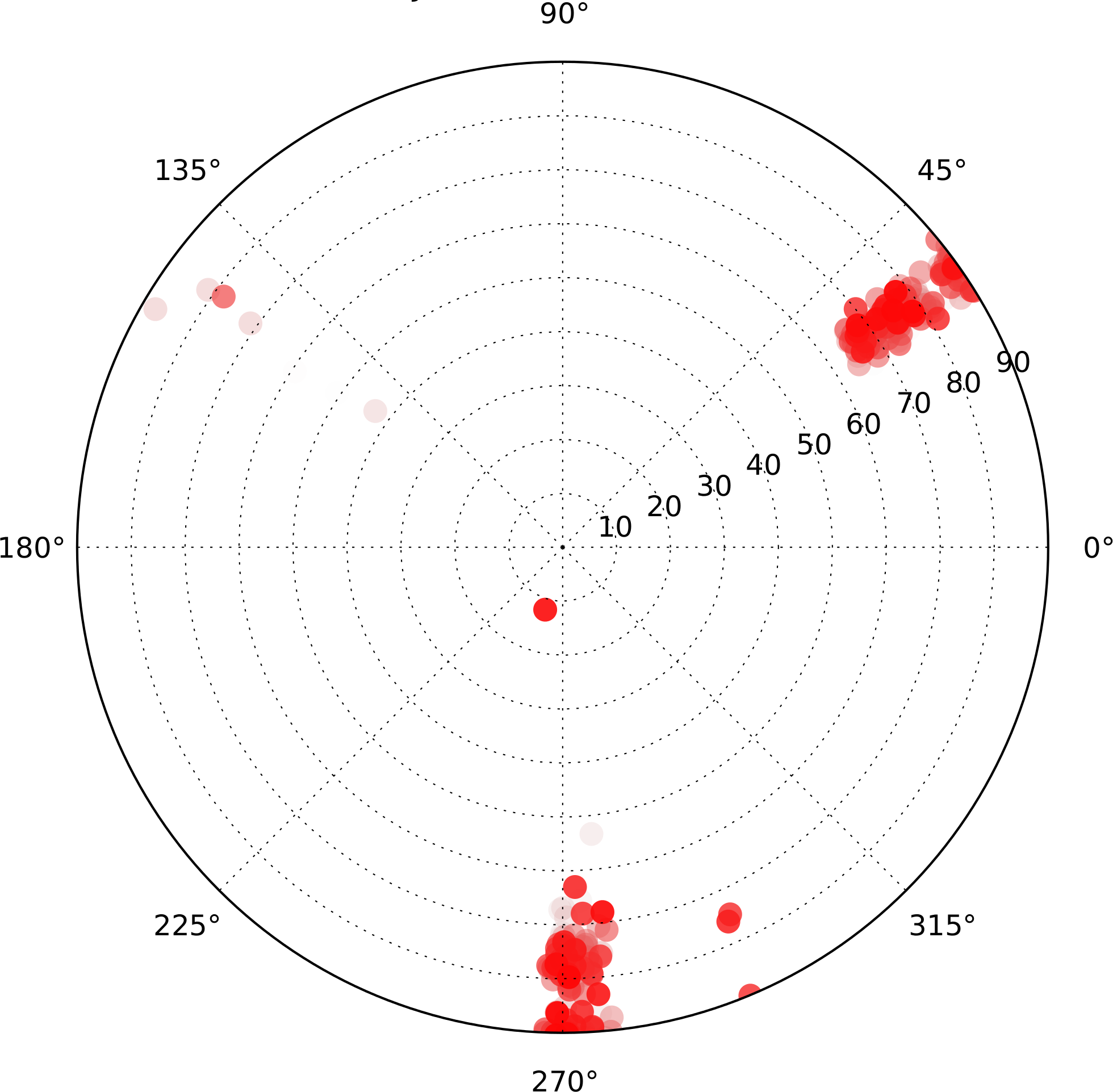}
 \end{center}
 \caption[Example of RFI localization at LOFAR]{Example of RFI localisation in early LOFAR data, showing several sources within an hour of observation. The direction reconstruction is not very accurate due to the fast method chosen, but sources are clearly identifiable. Figure courtesy A.~Corstanje.}
  \label{fig:lofar_rfi_localisation}
 \end{figure}

In early LOFAR data, it was straight-forward to identify near-field sources from something as simple as timing of amplitude threshold crossings. In \figref{fig:lofar_rfi_localisation} the results of a plane-wave fit to the times of amplitude threshold crossings are shown. Using the full buffered data would increase the resolution of the localisation, which however was not necessary in this case. Fitting a spherical wavefront provides additional information as it provides a distance estimate of the source (e.g.\ \cite{Bourke2017}).

The experience from LOFAR, as well as other experiments, is that little can be done to mitigate the effects of broad-band RFI once it is in the data, apart from discarding certain amounts of data where the contaminatory effect is severe. It is therefore important to identify and remove local sources of RFI. Such localisation can be difficult using beamformed data where RFI enters through the far sidelobes, and especially for sources that are in the near-field of the instrument. Buffered raw data allows for full flexibility in identifying the sources, especially when the sources are intermittent transients. An almost famous example were the electric fences surrounding the LOFAR central core, which were only visible as elevated noise in the beamformed data. Only when using the data from the transient buffers, the source could be pin-pointed as the very local phenomenon of electric fences \citep{Bourke2017}. It should be pointed out though, that the RF sources usually have to be somewhat close to obtain useful results of the spherical fit, due to their low elevation arrival directions and the signal propagation above ground \citep{Monroe2018}.

\subsection{OVRO-LWA -- Broadband RFI}

The OVRO-LWA in California \citep{2015Kocz} currently consists of about 250 dual-polarized antennas, spread over an roughly 200 meter wide core. Nearly the full band-width of data from 0 -- 100 MHz is collected at a central location, where they are typically correlated for astronomical observations. In a pilot-effort, the code on the FPGAs was replaced to allow for the real-time detection of cosmic-ray signals in the raw-data \citep{Monroe2018}. While there are no buffers involved that store the data for seconds, the on-board buffer on the FPGA acts as temporary storage until a trigger decision is formed. By having all data at a centralized location, there are few limitations on the data-rate of this read-out process. It should be noted that this was only possible as a dedicated effort and not in-parallel on ongoing astronomical observations.

\begin{figure}
 \begin{center}
  \includegraphics[width=0.9\linewidth]{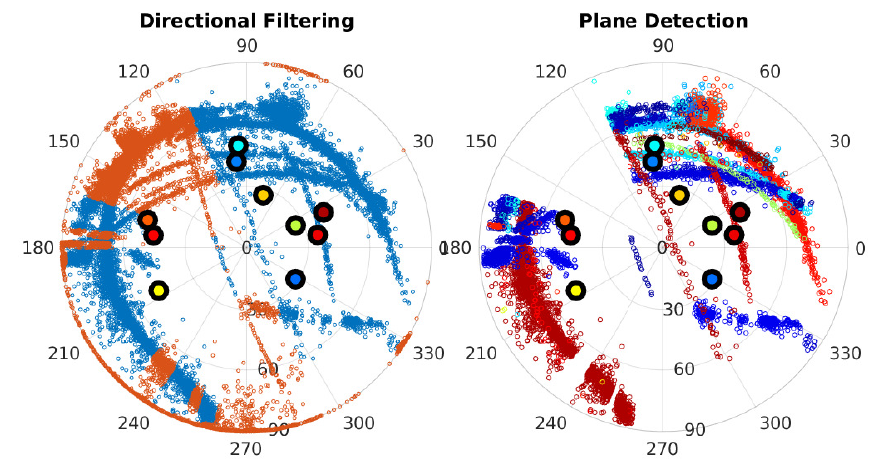}
 \end{center}
 \caption[Example of RFI localization at the OVRO-LWA]{Example of RFI localisation at the OVRO-LWA. The left figure shows all 473 thousand reconstructed directions of a dataset. Most events arrive from the orange regions in azimuth and are removed. The right figure shows the remaining directions and their timing within the observation is encoded in colour. Airplane track are nicely visible. The highlighted dots in the figure are cosmic ray candidates detected with the OVRO-LWA. See \cite{Monroe2018} for details on this analysis.}
  \label{fig:ovro_rfi_localisation}
 \end{figure}

These most recent efforts at the OVRO-LWA using the raw data have led to the identification of self-generated RF from the A/C unit of the signal processing shelter that were not found previously in imaging data \citep{Monroe2018}. Also, sparking and arching at local power-lines has been identified and reported to the utility company, which reported that they have no way to remotely identify such problems. For self-triggered cosmic ray searches the problematic directions have to be flagged (see Figure \ref{fig:ovro_rfi_localisation}), resulting in a loss in acceptance and lifetime. The successful localisation now allows for the mitigation of the RF source, which will also increase the data quality.
OVRO-LWA also detected transient pulses from airplanes, and one could envisage that buffered data could act as a monitoring tool for these and other regularly occurring RF sources, which would otherwise interfere with (in particular transient) observations.

\subsection{ATCA -- Broadband RFI}
\label{sec:rfi_lunaska_atca}

\begin{figure}
 \begin{center}
  \includegraphics[width=\linewidth]{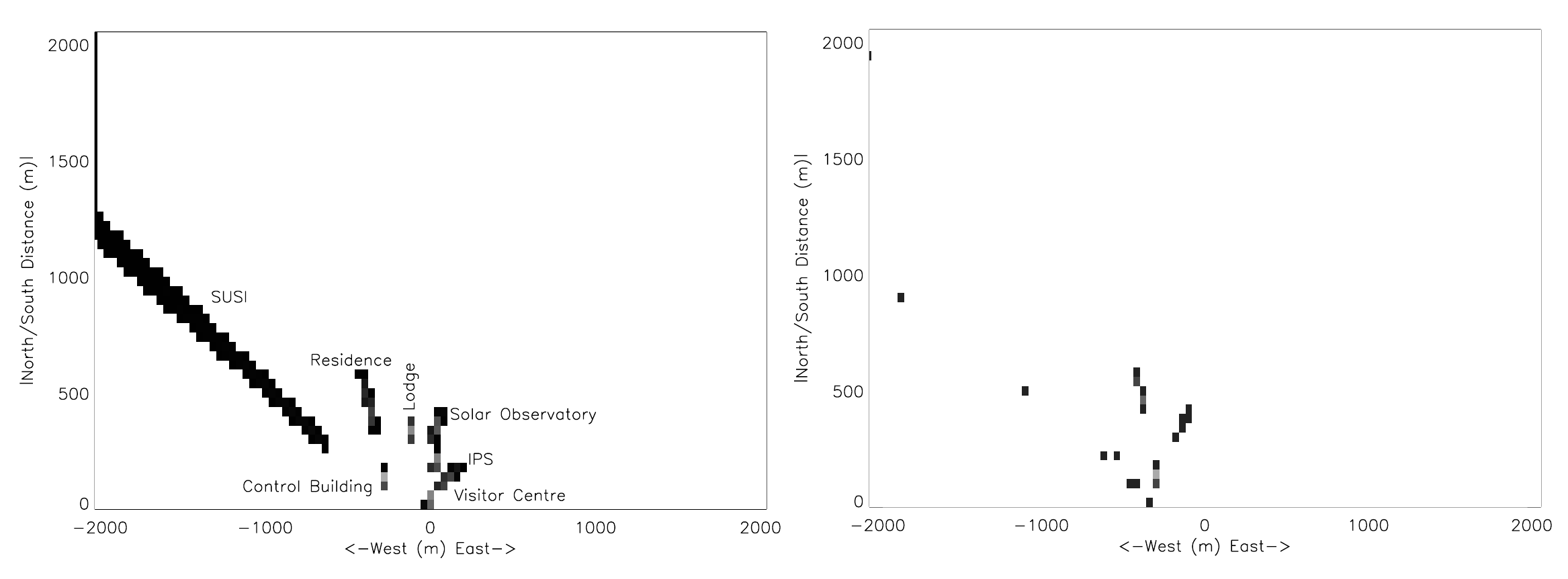}
 \end{center}
 \caption[LUNASKA/ATCA reconstruction of RFI origin locations (1)]{Expected (left) vs.\ observed (right) reconstructed positions of RFI generated by local structures at the ATCA site, as observed by the LUNASKA experiment for one period/telescope configuration \citep{cwjthesis}. The simulation results were obtained using the known locations of the site buildings and ATCA antennas to calculate impulse arrival times at each antenna. These were then smeared with an assumed uncertainty of $\pm 14.65$\,ns ($\pm30$\,samples), and reconstructed with a spherical wave solution. Clearly, the residence, lodge, and control building are detected in the observed positions. All positions on the South side have been mapped to the North side, since the East--West baseline could not distinguish between North and South.}
  \label{fig:atca_period_1}
 \end{figure}
 
\begin{figure}
 \begin{center}
 \includegraphics[width=1.0\linewidth]{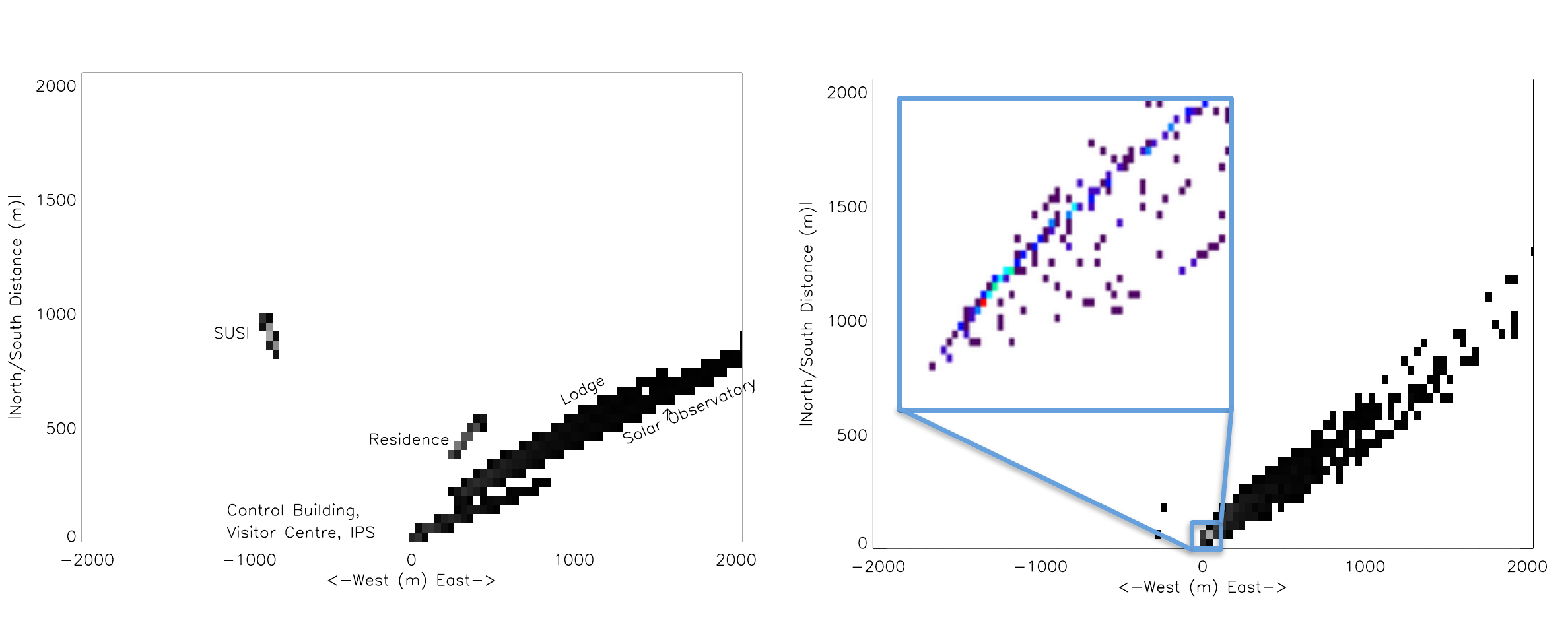}
 \end{center}
 \caption[LUNASKA/ATCA reconstruction of RFI origin locations (2)]{Expected (left) vs.\ observed (right) reconstructed positions of RFI generated by local structures at the ATCA site, as observed by the LUNASKA experiment for a different period/telescope configuration to that of Figure \ref{fig:atca_period_1}. The inset shows a zoom with each pixel representing a $2\times2$\,m resolution, clearly resolving the source of the RFI (red dot), which was a small solar telescope located at the South-East corner of the Visitors' Centre.}
  \label{fig:atca_period_2}
 \end{figure}

The LUNASKA experiment at the ATCA \citep{2010PhRvD..81d2003J} was the first experiment to use the Compact Array Broadband Backend (`CABB'; \cite{2011MNRAS.416..832W}), which had been installed on three antennas at the time. A peak-detection algorithm triggered each antenna individually whenever the voltage on either A or B polarisation reached above a tuneable threshold. For each trigger, $256$ 8-bit samples of 2,048\,GHz sampled baseband data was returned. A total of $\sim1.5\times10^7$ such candidates were recorded this way in $\sim$33\,hr of observations.

Impulsive events could be timed to within $0.5$\,ns, giving a spatial resolution of $\sim15$\,cm. However, because all three antennas lay on an East--West baseline, there was cylindrical degeneracy in wavefront solutions. Nonetheless, the locations of local sources of RFI could be identified, and matched well with the locations of buildings on site (see Figure \ref{fig:atca_period_1}). The strongest source of RFI identified (see Figure \ref{fig:atca_period_2}) was located at the South-East corner of the Visitors' Centre, and was later identified as a small solar telescope.

\begin{figure}
\begin{center}
\includegraphics[width=0.7\textwidth]{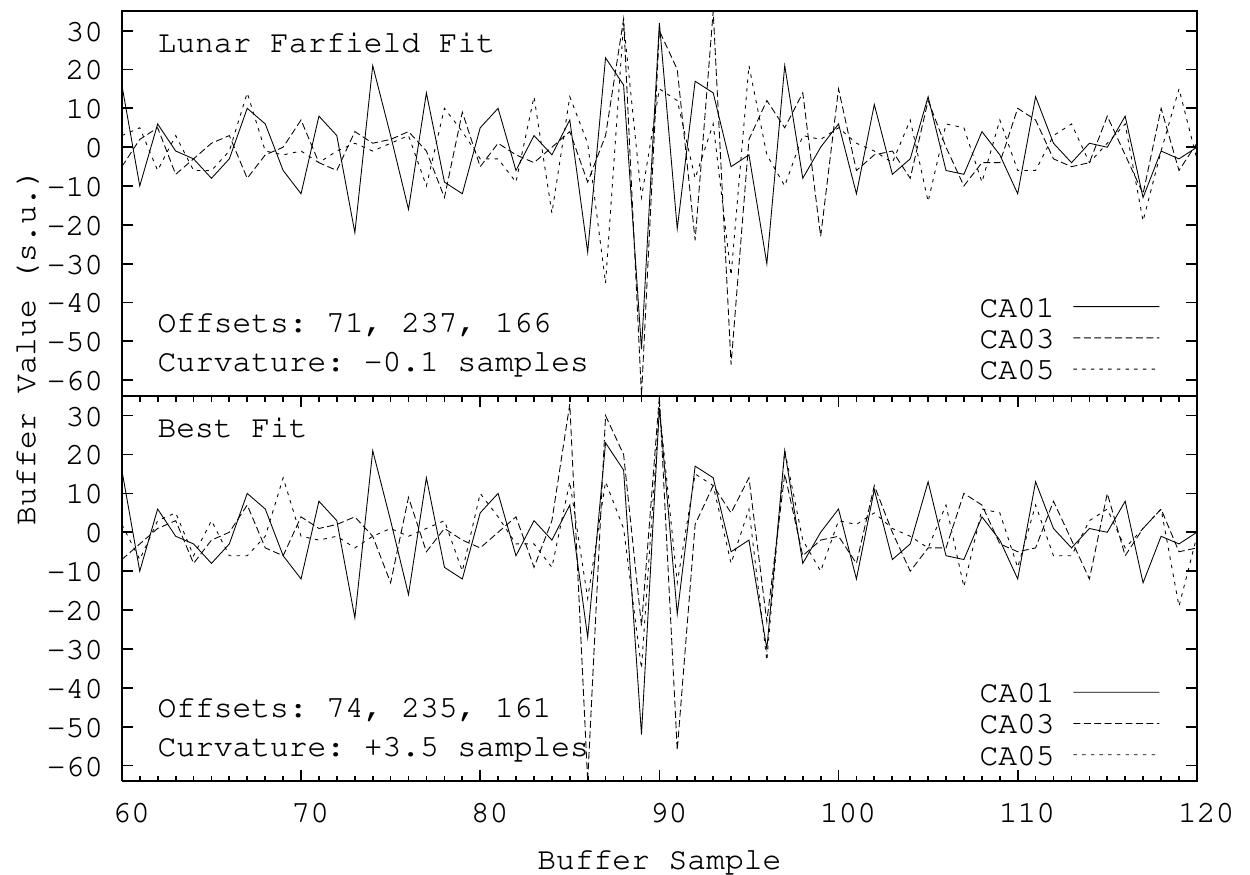}
\caption[Pulse timing alignment in baseband data]{Demonstration of the precision of high-time-resolution data. The best far-field (planar wavefront) fit (top) is compared to the best unrestricted (curved wavefront) fit (bottom) for a single polarisation B. The curvature measured, $3.5$\,samples, corresponds to a time offset of $1.7$\,ns, and a source distance of $\sim100$\,km.}
\label{fig:atca_candidate}
\end{center}
\end{figure}

For the RFI reconstructions shown above, only the raw trigger times (voltage above threshold) have been used, with no further processing, hence the large spread of reconstructed directions. When fully processed however, extreme precision becomes possible. Figure \ref{fig:atca_candidate} shows the only RFI event arriving with timing consistent with the direction of the Moon. By measuring the wavefront curvature, its nature as a nearfield event -- albeit at a distance of 100\,km -- is clearly identified.

\begin{figure}
\begin{center}
\includegraphics[width=\textwidth,trim={0 8.5cm 0 2cm},clip]{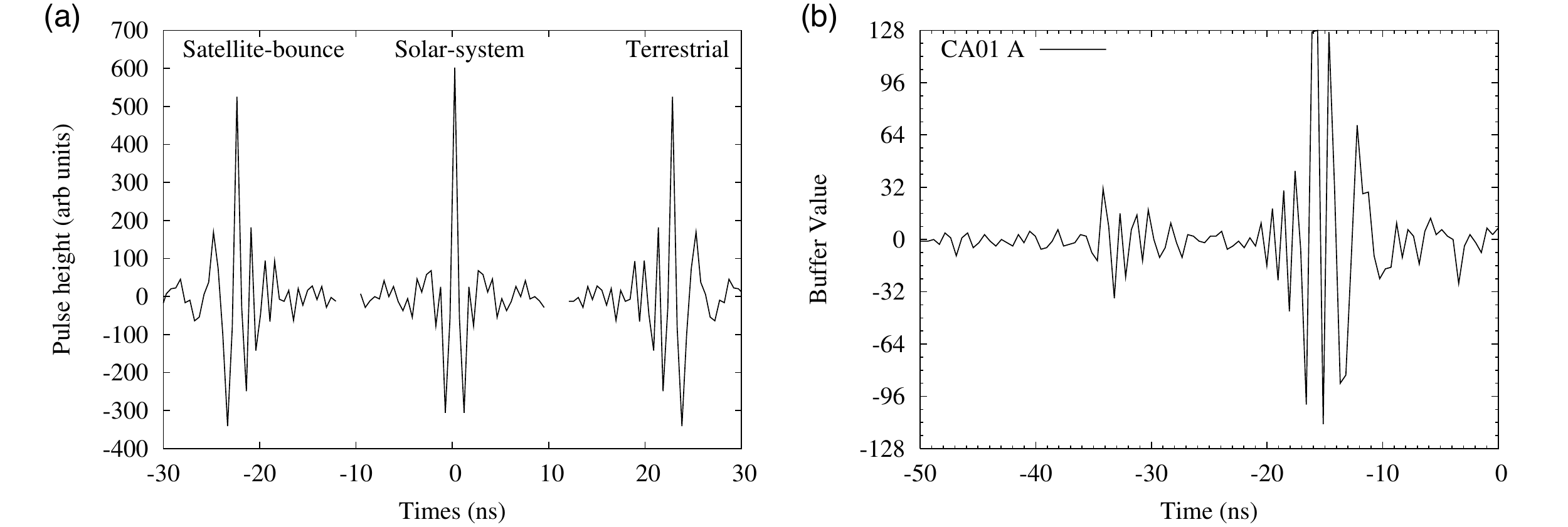}
\caption[Time-domain RFI from telescope calibration system]{Simulated (left) and detected (right) transient signals from the LUNASKA experiment at the ATCA. In both panels, larger times are earlier, i.e.\ time increases to the left. The simulated transients are all band-limited impulses of different origins. After passing through the dedispersion filter to compensate for the Earth's ionosphere, a pulse of solar-system (e.g.\ Lunar) origin will be correctly de-dispersed to a monopolar pulse. A pulse of terrestrial origin, having no original dispersion, will arrive with lower frequencies first; while a satellite-bounce pulse might pass through the ionosphere twice, and be only partially de-dispersed. Right: an actual pulse detected at the LUNASKA ATCA experiment, with primary pulse from -10 to -20\,ns, and secondary component from -25 to -35\,ns. The terrestrial nature of the primary pulse is clear by comparison to simulations. The source of the delayed secondary impulse was identified using the $\sim17$\,ns lag time, which corresponded to twice the length of the dedispersion filter, and was due to an impedance mismatch at its connection point.}
\label{fig:atca_noise_diode}
\end{center}
\end{figure}

The most common source of RFI however was generated locally in each antenna. This was the noise diode used to measure $T_{\rm sys}$, which produced regular impulses at its switching rate of $8$\,Hz. An example is given in Figure \ref{fig:atca_noise_diode}. Buffered data was used to both identify its nature, and locate the source of a reflection within the signal chain.

\section{Calibration}
In order to reconstruct measured air showers, an excellent model of the antenna characteristics, including the full bandpass of the system, is needed. The cosmic ray community has developed and tested many methods to calibrate antennas and the system response \citep[e.g][]{LOPES2008,Nelles2015,LOPES2016,Auger2017}.

There are two aspects of antenna calibration that are relevant: the directional sensitivity and the absolute calibration. 

For the directional sensitivity, ultimately a finite-element modelling of the antennas is needed, since the three-dimensional electric field vector has to be reconstructed for all arrival directions of the air shower. A dedicated calibration measurement can never cover all directions on a smooth grid, so some kind of modelling has to be involved that is then cross-checked with data. 
With astronomical observations, it is difficult to calibrate the details of the element beam (in the case of LOFAR) as either the sources are extended or they are not visible in the signal of a single element. This is different for cosmic ray signals, which is why they require the highest precision of understanding of the individual antennas. 

\begin{figure}
\begin{center}
\includegraphics[width=0.49\linewidth]{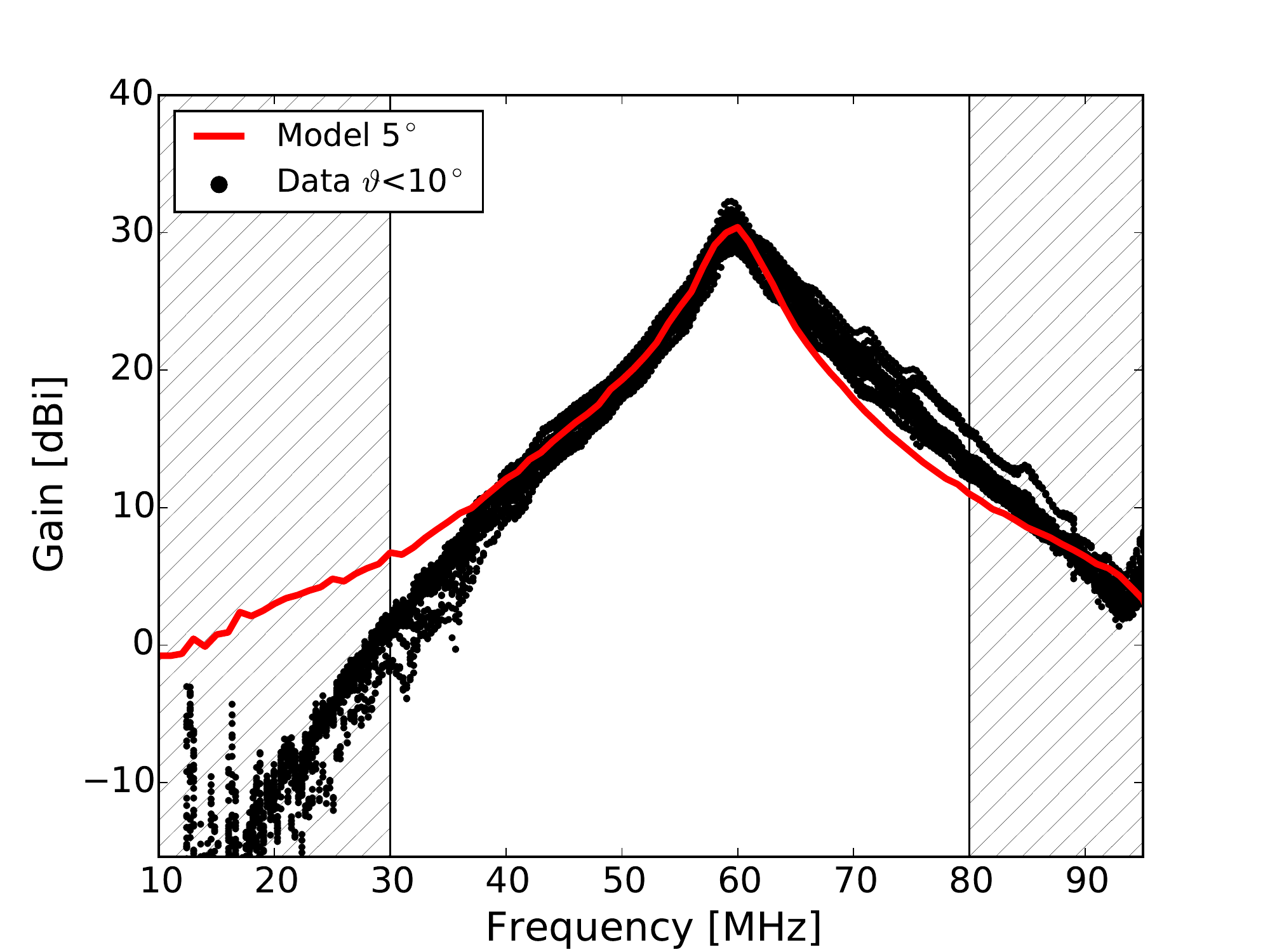}
\includegraphics[width=0.49\linewidth]{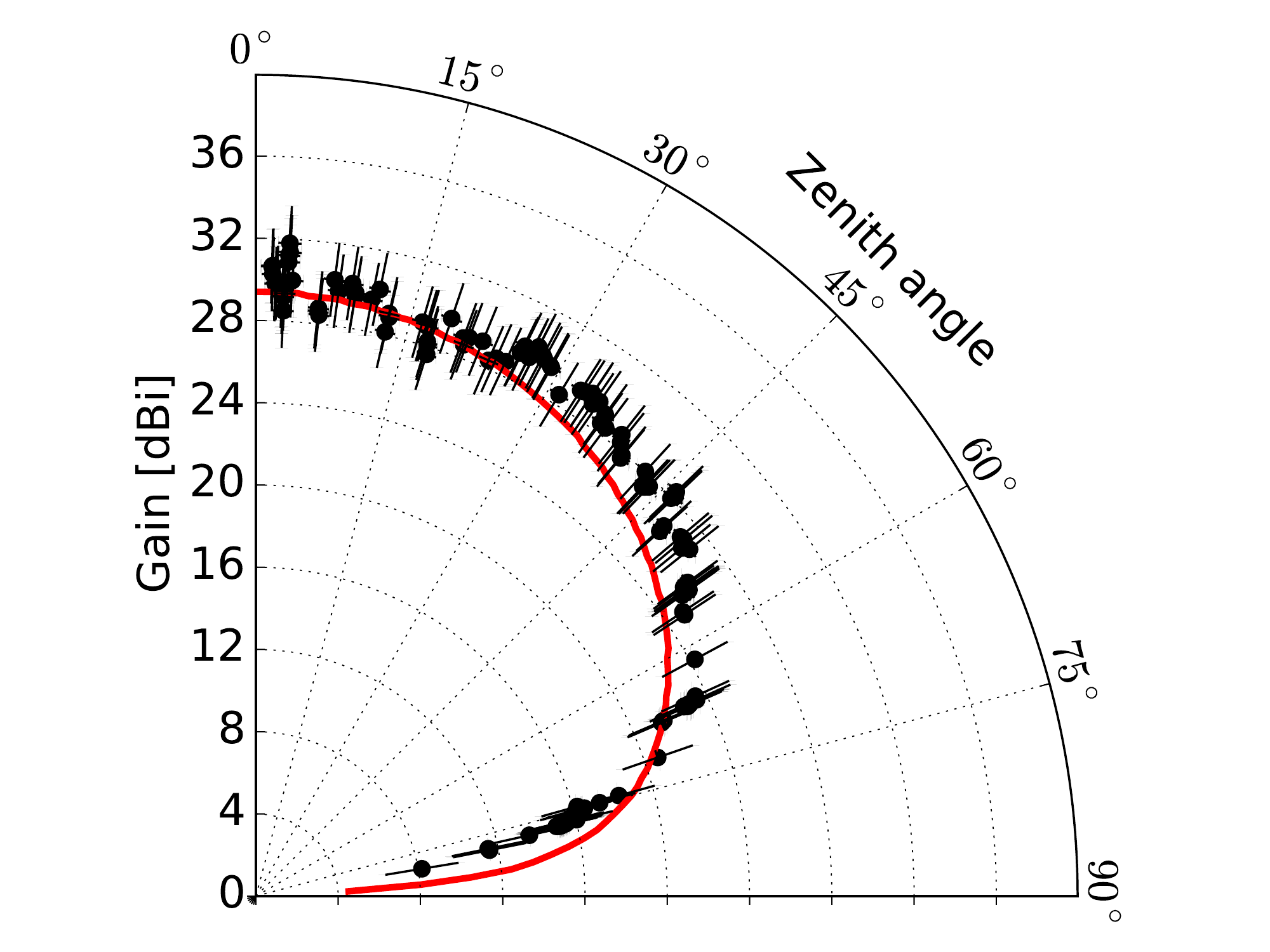}
\caption[Antenna model and octocopter measurements]{Absolute antenna gain as measured from an octocopter flight in comparison to an element beam model of the LOFAR LBA. Shown are both the frequency dependence (left) and the angular dependence (right) \citep{Nelles2015}.}
\end{center}
\label{fig:antenna_model}
\end{figure}

\begin{figure}
\begin{center}
\includegraphics[width=0.49\linewidth]{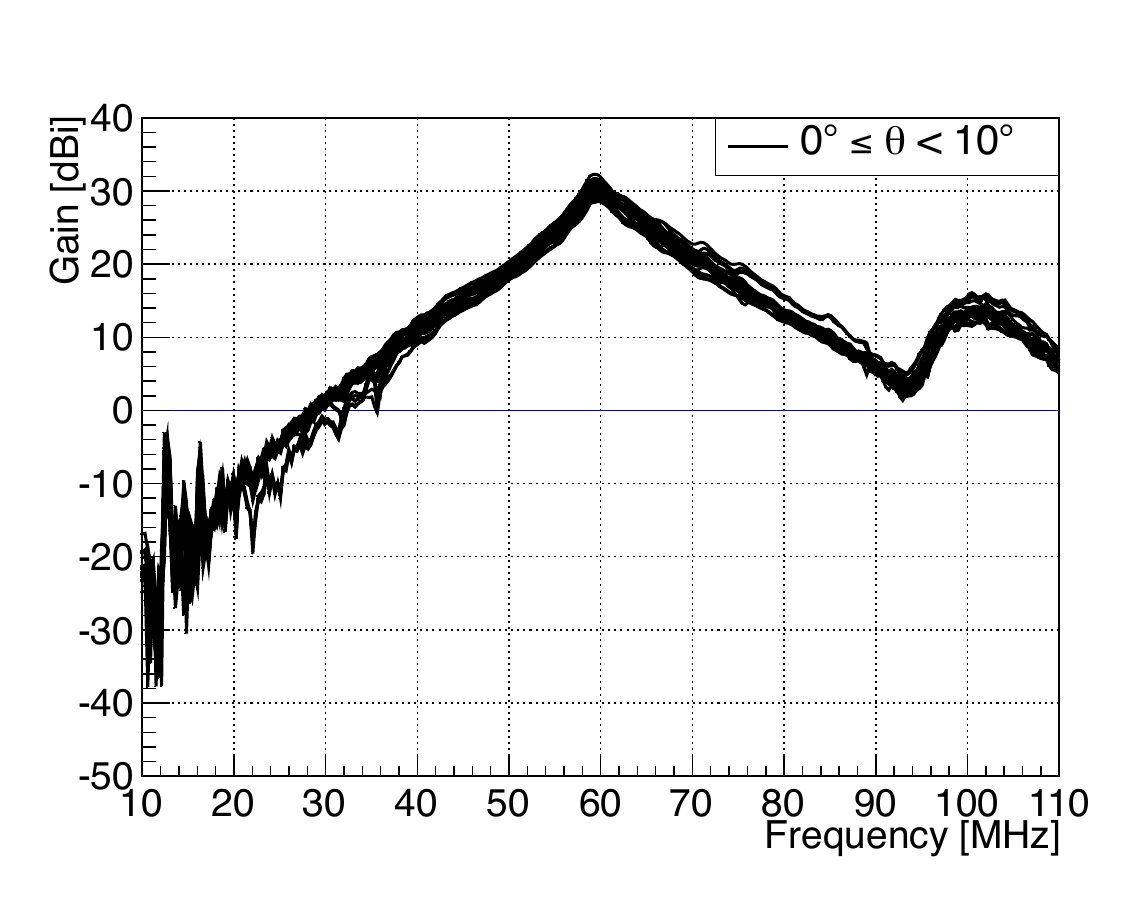}
\includegraphics[width=0.49\linewidth]{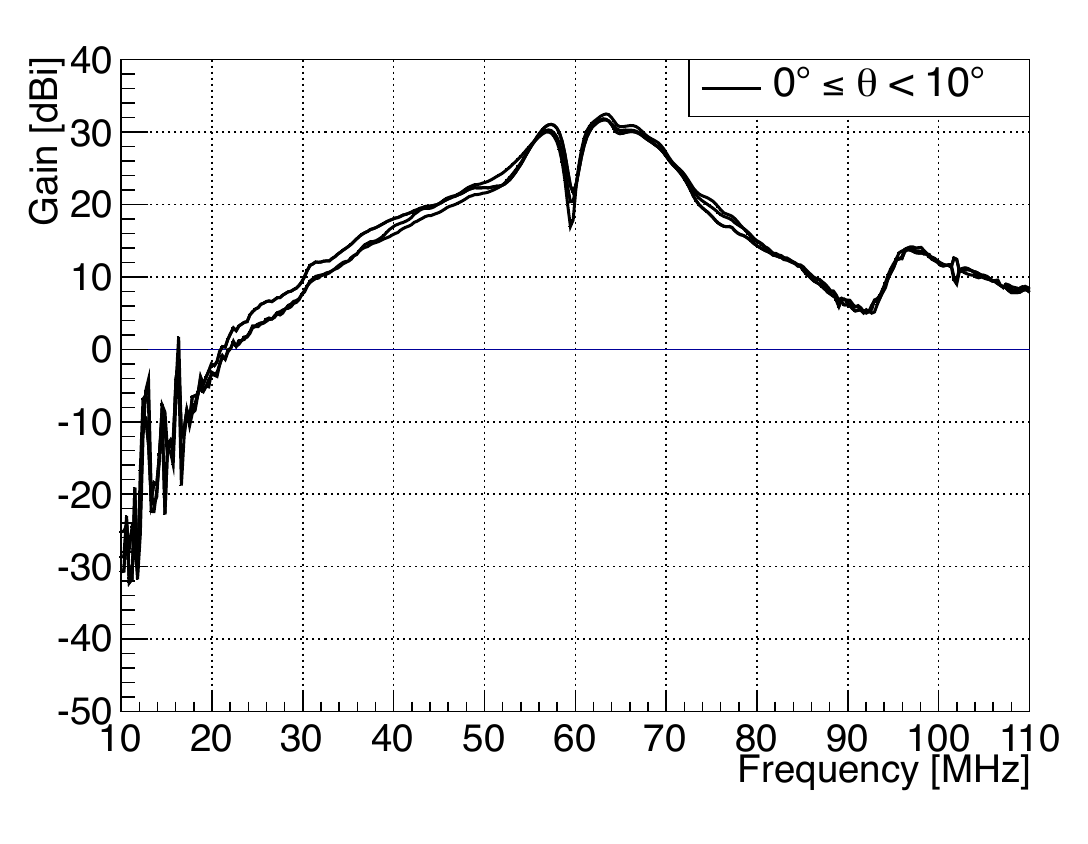}
\caption[Mutual coupling LOFAR LBA Inner]{Measurements of two LOFAR LBAs during an octocopter campaign \citep{Krause2013}. The left measurement was taken at one of the outermost LBAs of a LOFAR station, while the right one was taken at the innermost LBA. The effect of the dense antenna spacing on the shape of the bandpass is obvious.}
\label{fig:inner_dip}
\end{center}
\end{figure}

Several dedicated campaigns have been organized at LOFAR using reference calibration sources either mounted to a crane or to an octocopter drone \citep{Nelles2015}. Both calibration campaigns provided valuable information to improve the element models. They were also a good diagnostics tool showing, for example, that some LBAs are too close together so that the shape of their bandpasses change as shown in \figref{fig:inner_dip}. While there are few options to mitigate this, calibration campaigns necessary for cosmic ray studies provide vital checks of system health. The use of dipole-level raw data avoids several complications. For example, one could simply trigger the read-out of buffers while the octocopter was overhead and later correlate the recorded GPS position with the timestamps of the data. In this way, the entire system, including amplifiers and signal cables, could be calibrated. While this was successfully done for system timing at LOFAR \citep{Nelles2015}, the achievable uncertainty on the system bandpass was too large given the limited number of buffer read-outs allowed. Data was taken to this end, but not calibration was published. 

However, such calibration measurements are both time-consuming and require significant processing, and still only provide a snap-shot of the system. Therefore, the LOFAR LBAs are currently continuously absolutely calibrated to $\sim 15$\% accuracy using the diffuse Galactic synchrotron emission \citep{Nelles2015, Mulrey2018}. Interestingly, the radio signals measured from cosmic rays are now so well understood that they themselves could be used to absolutely calibrate the system \citep{Auger2016PRL}. First principle calculations predict the signal power to better than 5\% uncertainty. Such a calibration would be absolute and independent from astronomical source modeling. The calibration assigns absolutely calibrated values to the measured electric field, which also makes it independent of imaging or other processing software. Such a calibration could be continuously run at a central processing facility. 

\section{System health and identifying technical problems}

The access to the raw data and having to process it for analysis immediately reveals technical problems at the most fundamental level. It also allows for specific feedback on where exactly a problem is observed as the signals are not averaged or summed before analysis. 

\subsection{Timing accuracy}

One important aspect of system health is timing accuracy, which we treat in terms of three components: calibration of timing offsets in the system, timing stability, and the identification of glitches. All three can be measured through raw buffered data, which retains the sampling time resolution.

\begin{figure}
 \begin{center}
  \includegraphics[width=0.55\linewidth]{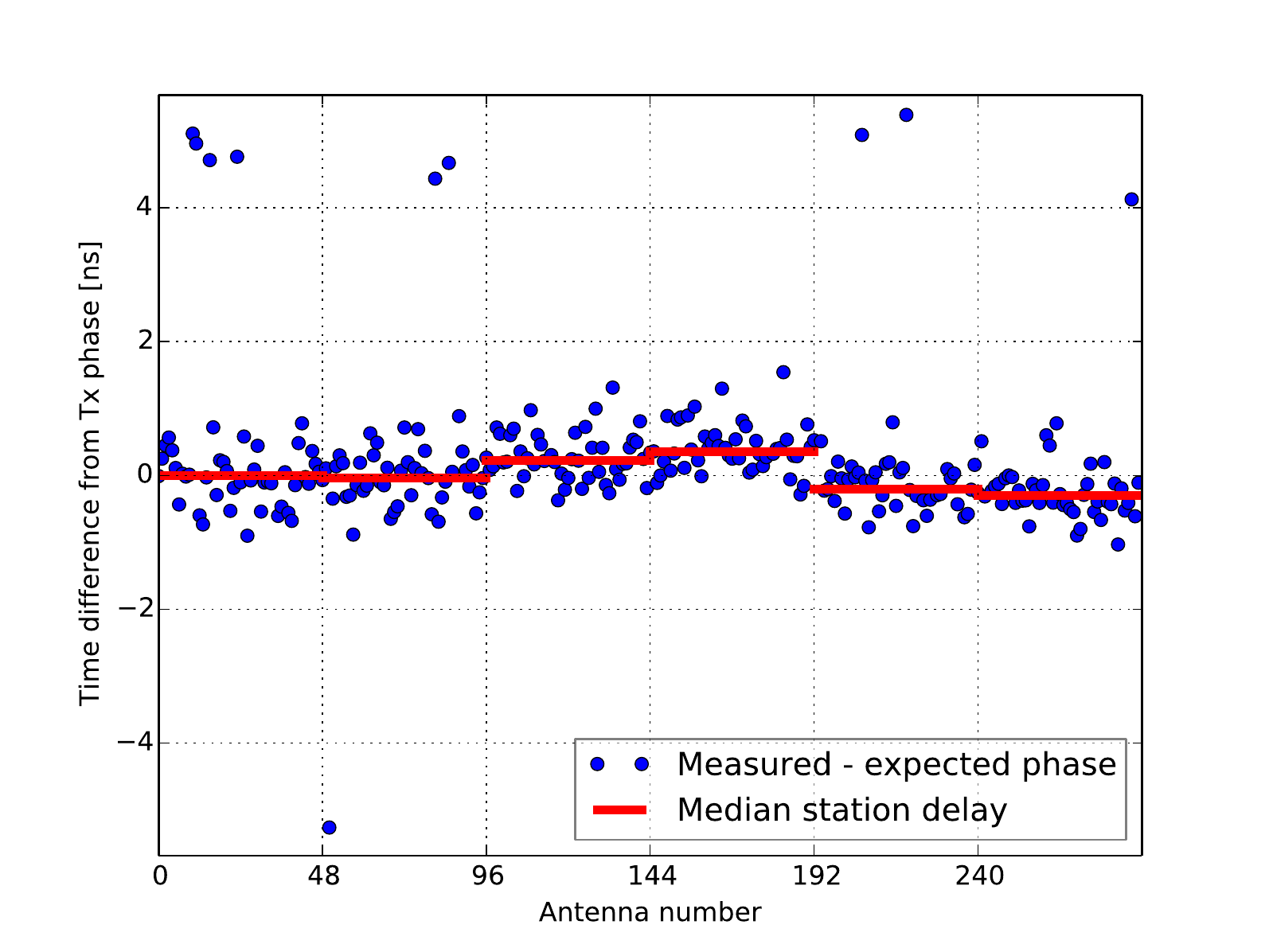}
  \includegraphics[width=0.49\linewidth]{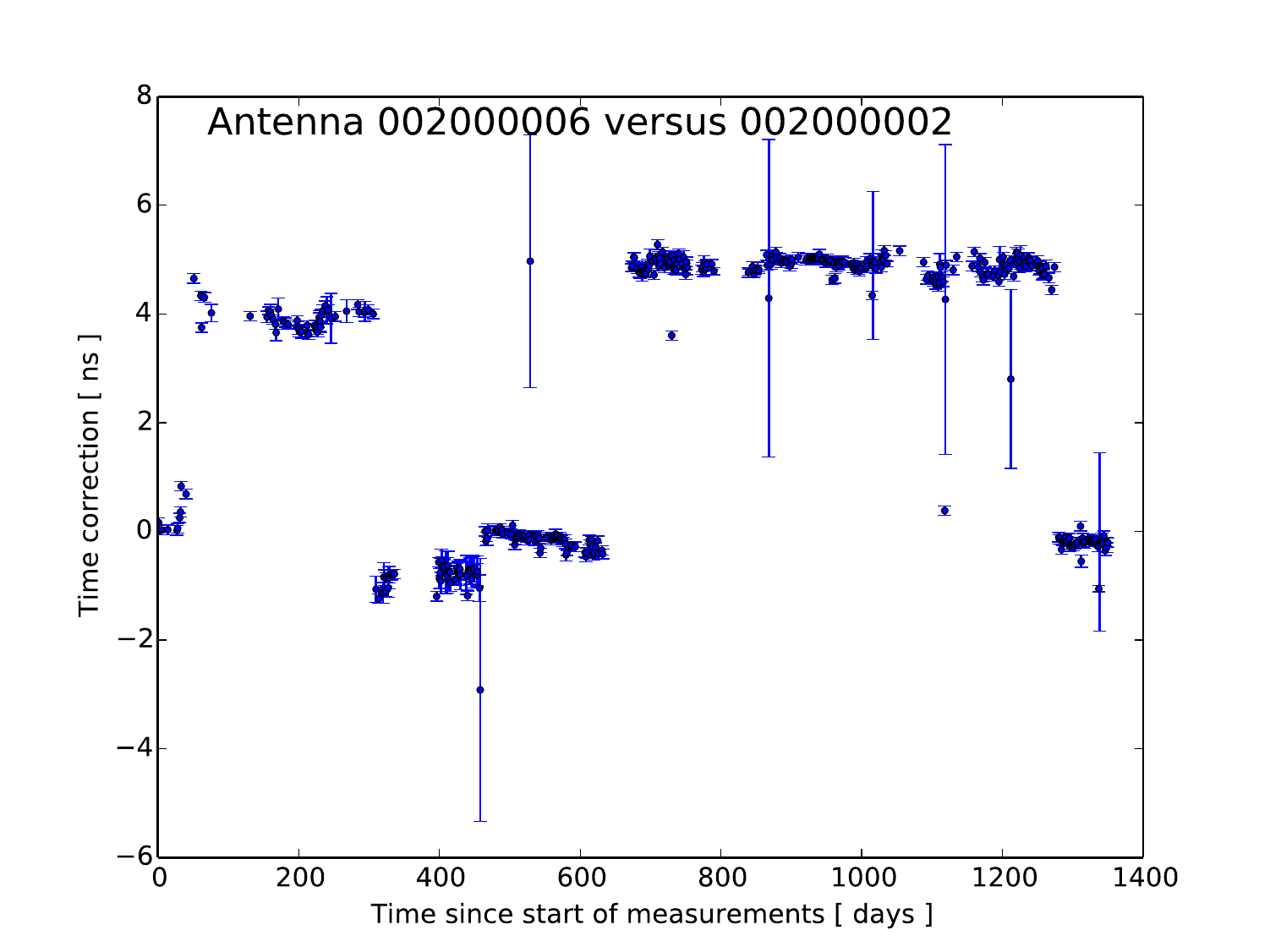}
  \includegraphics[width=0.49\linewidth]{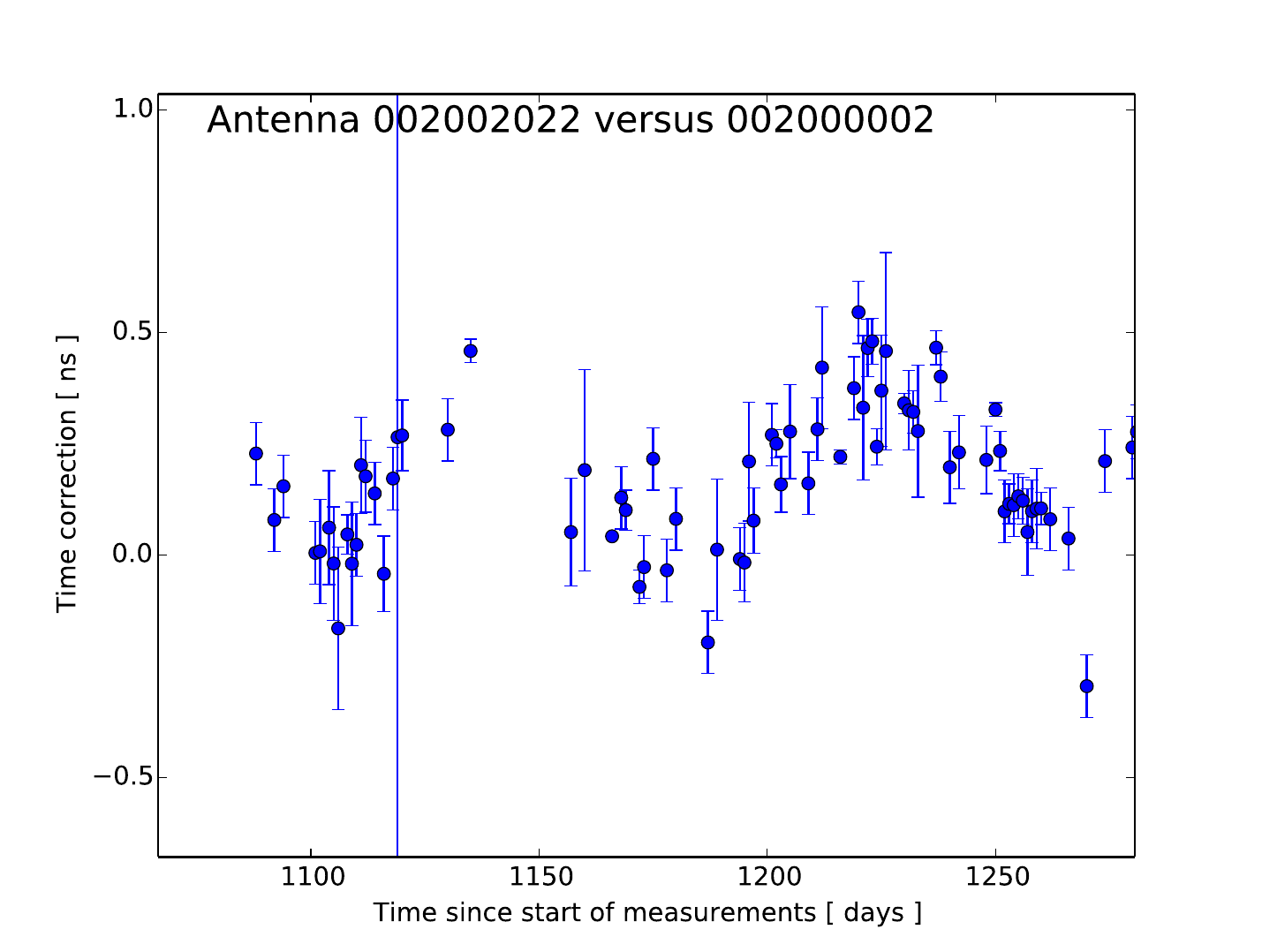}
 \end{center}
 \caption[Clock errors in early LOFAR data]{Top: Intrinsic time-delay per antenna of all LOFAR LBA stations on the central core (superterp) as derived from an RF transmitter. Shown is both the delay per antenna and the median station delay. In this case, all median delays are consistent with zero, so all system delays have been accounted for. 
Bottom: Phase difference of one pair of antennas as function of time for a large fraction of cosmic ray data (2011-2014) and a zoom to a period of 200 days. Both the 5 ns clock jumps, as well as a slow drift of the clocks can be observed.
Note that these figures show early LOFAR data, and the 5 ns clock jumps have been fixed, also following reports from the CR key science project.}
  \label{fig:clock_erros}
 \end{figure}

As shown in \cite{Corstanje2016}, certain continuously present narrowband RF lines (such as FM stations, even though heavily suppressed by the bandpass in LOFAR) can be used to track the relative phase stability of the system. \figref{fig:clock_erros} shows three examples from LOFAR. From the raw data obtained after cosmic ray triggers, the phase difference of RF lines between antenna pairs is calculated. Irrespective of the transmitter behaviour, this phase difference between antenna pairs should be stable as function time, since it is due to the spatial offset between the two antennas. The method is able to identify constant delays between stations (i.e.\ cable delays that had not fully been accounted for), clock glitches (200 MHz sampling = 5 nanosecond glitch) and clock drifts to better than the sampling frequency. All the necessary individual measurements contained in these plots are standard products of the cosmic ray pipeline at LOFAR. 

\begin{figure}
 \begin{center}
  \includegraphics[width=0.7\linewidth]{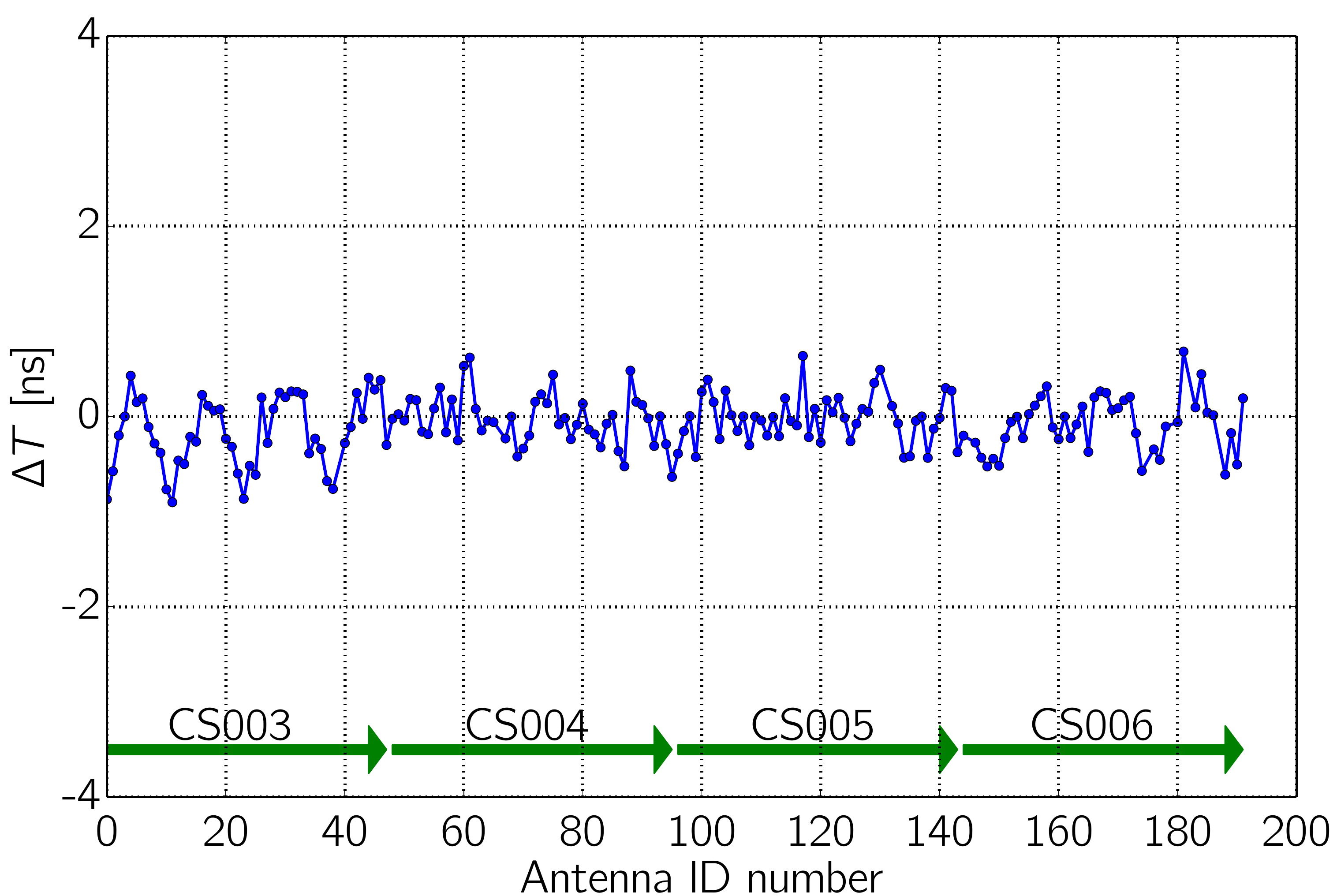}
 \end{center}
 \caption[Timing from octocopter measurement]{Timing difference between the expectation from the position of the octocopter and reconstructed arrival times per antenna \citep{Corstanje2016}. Overall, no large deviating times were observed. The residual oscillation in the time difference illustrates one difficulty of drone calibrations: The accuracy in positioning is challenging and uncertainties lead to a mismatch of reconstructed and measured position, which limits the accuracy of the resulting timing calibration to 0.5 ns accuracy.}
  \label{fig:oct-time}
 \end{figure}

A similar RF line analysis has been conducted at the Pierre Auger Observatory using a dedicated permanently installed transmitter, which is of course not a viable option near a radio telescope \citep{Auger2016}. Alternatively, one can use octocopter calibrations (see \figref{fig:oct-time}) or airplane signals \citep{Auger2016}, however, they only allow for snapshots and require additional effort in setting up the measurements campaigns and/or the airplane tracking.

\subsection{Non-linearity}

As cosmic-ray events involve an abrupt, high-magnitude excursion in the electric field measured by an antenna, measuring them depends on the linear performance of amplifiers and other signal-path components out to voltages significantly beyond the thermal noise level.  Antenna-level buffers used to detect cosmic-ray events therefore provide a direct means -- and the motivated researchers -- to identify and diagnose non-linear behaviour.

A problem of this type was identified in the Parkes 21~cm multibeam through the use of baseband buffering for the LUNASKA experiment \citep{2015APh....65...22B}.  This experiment analysed dual-polarisation data from four beams of the receiver, sampling the downconverted signal at 8~bits and 1.024\,GHz. Data were passed through a FIR filter to account for ionospheric dispersion; raw and filtered data were buffered for 4\,$\mu$s, and read out when filtered voltage magnitudes exceeded a pre-defined threshold.

\begin{figure}
  \begin{center}
   \includegraphics[width=0.7\linewidth]{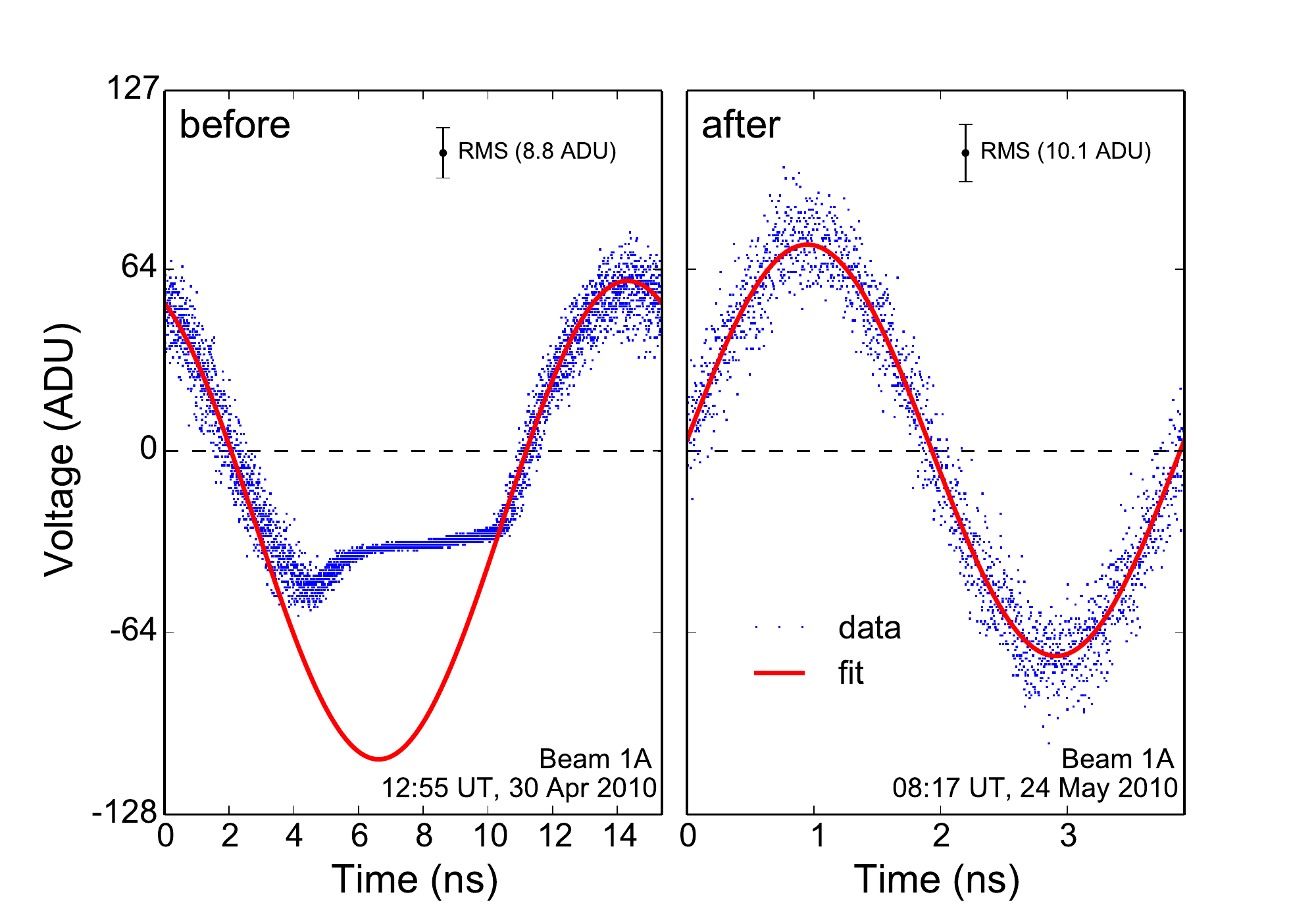}
  \end{center}
  \caption[Non-linear behaviour identified in Parkes receiver]{Single buffers of sampled values, recorded while transmitting a sine wave signal directly into the receiver, folded to the period of the signal.  Left and right panels show data from before and after additional attenuation was inserted early in the signal path to restore linear behaviour at large voltage amplitudes.  The solid line shows a sine-wave fit to the data near zero voltage, where the signal behaves linearly in both cases.}
  \label{fig:sinetest}
 \end{figure}

While all receivers maintained good linearity for positive peaks in the voltage, negative peaks were substantially suppressed in one beam (see figure~\ref{fig:sinetest}).  This issue was identified while searching for cosmic-ray events, and diagnosed by the observers using a modified version of their bespoke calibration system.  It was later corrected by inserting additional attenuation before the component responsible for the non-linear behaviour (and additional amplification after this point).

\subsection{Hardware errors and faults}
 
\begin{figure}
 \begin{center}
  \includegraphics[width=0.7\linewidth]{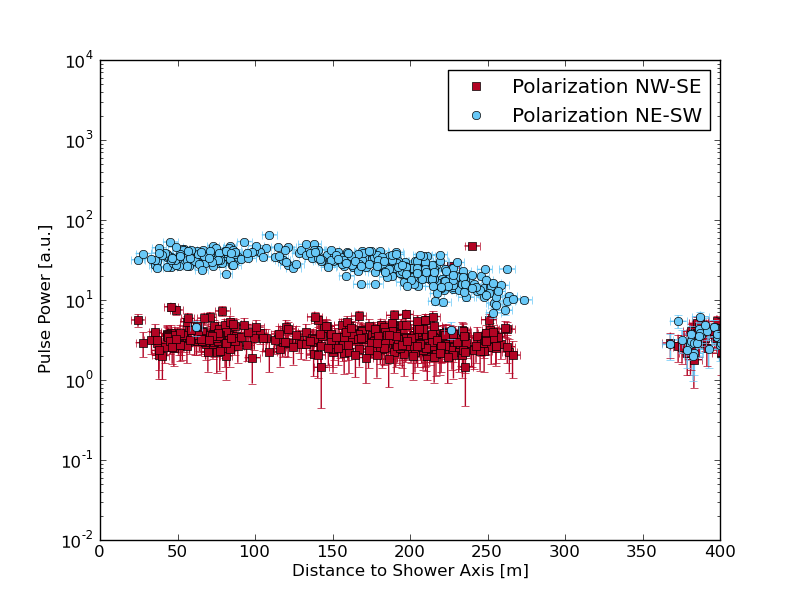}
 \end{center}
 \caption[Example of a swapped antenna polarization at LOFAR]{Single air shower recorded with three LOFAR LBA stations on the superterp. Shown are the raw data from both dipoles (red squares and blue circles). At a distance of about 60 meters as well as 220 meters, two blue points are visible in the red band of data, which belong to swapped polarizations. The red points in the blue bands are not visible, as they are plotted below the blue band. The red dot above the blue band is a broken channel.}
  \label{fig:swapped_pol}
 \end{figure}

Due to the strongly polarized nature of the cosmic ray signal, technical issues such as a swapped polarization within one antenna become immediately obvious, as shown in \figref{fig:swapped_pol}. In 2012, the Cosmic Ray Key Science Project reported 7 suspicious pairs of elements on the LOFAR superterp (of 480 pairs in total) derived from one year of cosmic ray data recorded at that point in time. Upon investigation, hardware issues with all of the dipoles were found, such as swapped coax cables at different patch panels, and misprinted amplifier PCBs (manufacturer failures). 

\begin{figure}
 \begin{center}
  \includegraphics[width=0.7\linewidth]{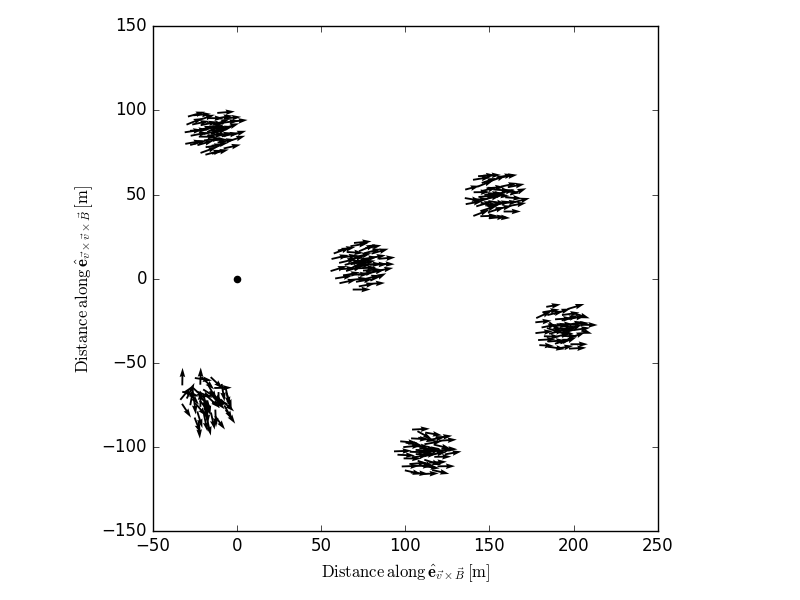}
 \end{center}
 \caption[Example of hardware issue influencing polarization]{Electric field vectors as reconstructed from data from LOFAR LBA inner stations for one air shower. In this early LOFAR data, CS004 (bottom left) shows an outlying behaviour from the expected polarization orientation for all stations along the x-axis. The axis is defined through the arrival direction of the cosmic ray and the magnetic field. Since arrival directions are random, many intrinsic polarizations are tested in this way.}
  \label{fig:pol_issue}
 \end{figure}

Also, another hardware problem has been identified via the polarization signature. The polarization of an air shower signal should be strongly polarized in the direction of the cross-product of the shower axis $\vec{v}$ and magnetic field $\vec{B}$ (x-axis of graph in Figure \ref{fig:pol_issue}). One LOFAR LBA inner station consistently showed a randomized polarization behavior, raising the suspicion that the timing was not stable. After an inspection of the station, the situation improved, which might be indicative of loose cables or a similarly transient problem. The precise reason for the original problem has never been established -- nonetheless, the use of buffered data helped to identify and fix it.

\begin{figure}
 \begin{center}
  \includegraphics[width=0.49\linewidth]{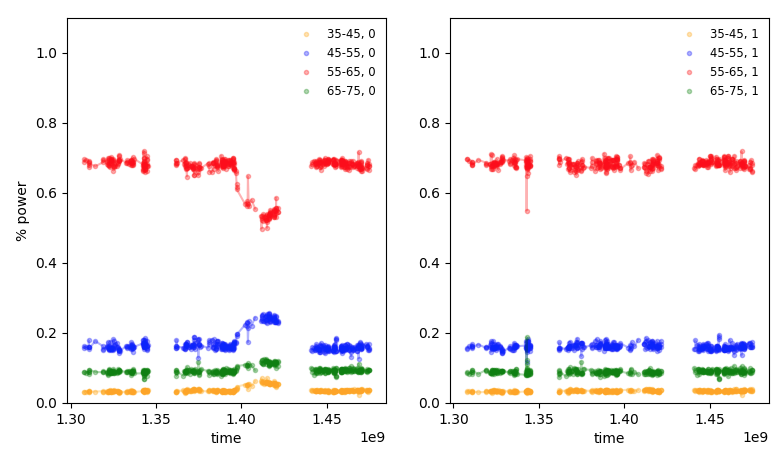}
  \includegraphics[width=0.49\linewidth]{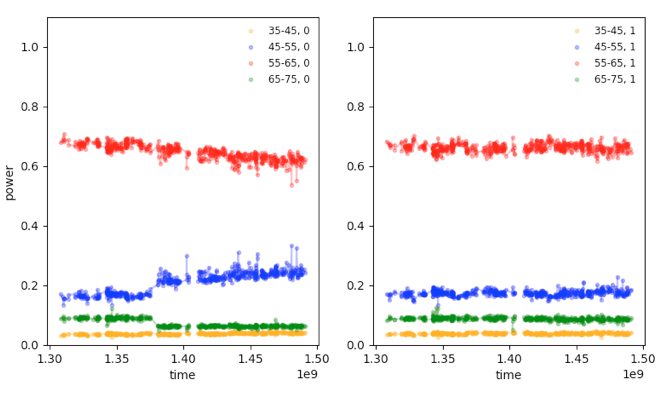}
 \end{center}
 \caption[Example of long-term study of antenna performance]{Fraction of power as function of time (in seconds) for different frequency bands (in MHz) in two polarization of one LBA antenna. Shown are two examples (two figures each) of badly behaving antennas. In the ideal case all lines should be perfectly horizontal, representing constant power. On the left an antenna showed multiple mechanical problems and was subsequently turned off twice. The right example shows a trend in one of the two dipoles. The latter behaviour is only visible in long-term monitoring. Figure courtesy K.~Mulrey.}
  \label{fig:long-term}
 \end{figure}

The recording of buffered baseband data for cosmic ray studies provides semi-regular snapshots of the entire telescope system, effectively providing a database for tracking systems stability with time.
As shown in \figref{fig:long-term}, hardware failures at the antenna level are easily detectable that might otherwise go unnoticed.
The method is particularly sensitive to long-term changes in the performance, as a cosmic ray data set needs to be comparable throughout in entirety. 
 
In addition to the items reported above, the LOFAR experience has shown that items like broken amplifiers and broken cables (additional time delay) were easily identified in cosmic ray data. Also, it should be noted that the LOFAR HBAs, which are pre-beamformed before being buffered, are much harder to monitor with buffered data and an absolute calibration has been prohibitively complicated for the CRKSP.

\section{Discussion}
The different examples provided above show that buffers are an easy tool to obtain both random and triggered diagnostics for system health and instrument performance. The data is typically lightweight, and dedicated monitoring tools do not require significant computing or high-level data processing skills. In an era where data volumes and instruments grow more and more complex, systematic low-level monitoring increases in importance to ensure a fast turn-around for maintenance and failure detection. 

The experience with LOFAR has shown that typical buffer users (members of the Cosmic Rays Key Science Project, the equivalent of the Focus Group: High Energy Cosmic Particles) automatically provide a variety of low-level monitoring as by-product of their analyses. It is interesting to note that in astroparticle physics, there is very little separation between engineering teams who build experiments, and physicists who derive high-level science outputs from them. This is mostly due to the fact that particle physics experiments are continuously pushing the understanding of elementary physics, which is why every malfunction of the system needs to be excluded as possible cause of unexpected behaviour. 

This mind-set is in contrast to the open-observatory model, whereby a telescope is made accessible to the worldwide community. This model has huge and undeniable benefits -- however, one downside is that it can lead to a separation between science and engineering teams.
The use of antenna-level buffering for cosmic ray studies would help to keep a close association between experimentalists and engineers.

It will be a rather straightforward process to implement findings of the cosmic ray pipelines in a centrally run database for system purposes. In fact, the LOFAR cosmic ray database already monitors many quantities on an at least daily basis, and has done so since 2011 through to the present. While being very successful, there are still lessons to be learned from the LOFAR experience.
In particular, the LOFAR cosmic ray pipeline never ran at centralized observatory (ASTRON) facilities. This increased the turn-around time on the reporting of malfunctions and the doubling of effort, as a parallel monitoring structure had been developed. This can be avoided in the case of the SKA by an early cooperation between experimentalists and engineers, which will be mutually beneficial for instrument stability and science return. Such direct co-operation was achieved for the experiments at ATCA and Parkes, where for the examples shown in Figures \ref{fig:atca_noise_diode} and \ref{fig:sinetest}, the delay between problem observation, identification, and solution was a few hours.

Furthermore, the results presented here from LOFAR, ATCA, Parkes, and OVRO-LWA resulted from studies aimed at detecting cosmic rays. This meant that once sufficient fidelity for doing so was achieved, the analyses stopped. It is expected therefore that dedicated studies using such data specifically for system monitoring will be able to produce results far beyond those shown here.

\section{Conclusion}

The use of antenna level buffers of time-domain voltage data are critical for the detection of cosmic ray pulses. However, the low-level information they provide -- and the flexibility in subsequent data processing -- also provides unique and critical information for RFI studies, identifying malfunctioning hardware, testing beam patterns, and monitoring system health. We have presented numerous examples of this from experiments at the ATCA, Parkes, OVRO-LWA, and LOFAR. The experience of the Cosmic Rays Key Science Project with LOFAR is most directly applicable to that of SKA1-LOW, both because of the similarity of the instruments, and because the utility of access to low-level system information will only increase with instrumental complexity.

The study of cosmic rays is also now sufficiently advanced that the sub-microsecond pulses generated by cosmic-ray-induced cascades can be accurately predicted with simulations. The study of cosmic rays with a radio telescope equipped with antenna-level buffers thereby provides a free calibration signal for a wide range of telescope parameters independent of astronomical source modeling.

\section{Acknowledgements}
We thank the LOFAR Cosmic Rays Key Science Project, the Lunar UHE Neutrino Astrophysics with the Square Kilometre Array (LUNASKA) collaboration, and our colleagues at the Owens Valley Long Wavelength Array (OVRO-LWA) for discussions and providing previously unpublished material for this summary of experiences with buffers at radio telescopes.  

\clearpage
\addcontentsline{toc}{section}{References}
\bibliography{refs}

\begin{thebibliography}{22}
\expandafter\ifx\csname natexlab\endcsname\relax\def\natexlab#1{#1}\fi

\bibitem[{{Aab} {et~al}\mbox{.}(2016){Aab}, {Abreu}, {Aglietta}, {Ahn}, {Al
  Samarai}, {Albuquerque}, {Allekotte}, {Allison}, {Almela}, {Alvarez
  Castillo}, \& et~al.}]{Auger2016PRL}
{Aab} A. {et~al.}, 2016, Physical Review Letters, 116, 241101

\bibitem[{{Aab} {et~al}\mbox{.}(2017){Aab}, {Abreu}, {Aglietta}, {Samarai},
  {Albuquerque}, {Allekotte}, {Almela}, {Alvarez Castillo},
  {Alvarez-Mu{\~n}iz}, {Anastasi}, \& et~al.}]{Auger2017}
{Aab} A. {et~al.}, 2017, Journal of Instrumentation, 12, T10005

\bibitem[{{Apel} {et~al}\mbox{.}(2016){Apel}, {Arteaga-Vel{\'a}zquez},
  {B{\"a}hren}, {Bekk}, {Bertaina}, {Biermann}, {Bl{\"u}mer}, {Bozdog},
  {Brancus}, {Cantoni}, {Chiavassa}, {Daumiller}, {de Souza}, {Di Pierro},
  {Doll}, {Engel}, {Falcke}, {Fuchs}, {Gemmeke}, {Grupen}, {Haungs}, {Heck},
  {Hiller}, {H{\"o}randel}, {Horneffer}, {Huber}, {Huege}, {Isar}, {Kampert},
  {Kang}, {Kr{\"o}mer}, {Kuijpers}, {Link}, {{\L}uczak}, {Ludwig}, {Mathes},
  {Melissas}, {Morello}, {Nehls}, {Oehlschl{\"a}ger}, {Palmieri}, {Pierog},
  {Rautenberg}, {Rebel}, {Roth}, {R{\"u}hle}, {Saftoiu}, {Schieler}, {Schmidt},
  {Schoo}, {Schr{\"o}der}, {Sima}, {Toma}, {Trinchero}, {Weindl}, {Wochele},
  {Zabierowski}, \& {Zensus}}]{LOPES2016}
{Apel} W.~D. {et~al.}, 2016, Astroparticle Physics, 75, 72

\bibitem[{Bourke(2017)}]{Bourke2017}
Bourke S., 2017,
  http://www.astron.nl/ilttom2017/Documents/ILTTOM2017\_2\_SE607\_3.pdf

\bibitem[{{Bray} {et~al}\mbox{.}(2015){Bray}, {Ekers}, {Roberts}, {Reynolds},
  {James}, {Phillips}, {Protheroe}, {McFadden}, \&
  {Aartsen}}]{2015APh....65...22B}
{Bray} J.~D. {et~al.}, 2015, Astroparticle Physics, 65, 22

\bibitem[{{Buitink} {et~al}\mbox{.}(2016){Buitink}, {Corstanje}, {Falcke},
  {H{\"o}randel}, {Huege}, {Nelles}, {Rachen}, {Rossetto}, {Schellart},
  {Scholten}, {Ter Veen}, {Thoudam}, {Trinh}, {Anderson}, {Asgekar}, {Avruch},
  {Bell}, {Bentum}, {Bernardi}, {Best}, {Bonafede}, {Breitling}, {Broderick},
  {Brouw}, {Br{\"u}ggen}, {Butcher}, {Carbone}, {Ciardi}, {Conway}, {de
  Gasperin}, {de Geus}, {Deller}, {Dettmar}, {van Diepen}, {Duscha},
  {Eisl{\"o}ffel}, {Engels}, {Enriquez}, {Fallows}, {Fender}, {Ferrari},
  {Frieswijk}, {Garrett}, {Grie{\ss}meier}, {Gunst}, {van Haarlem}, {Hassall},
  {Heald}, {Hessels}, {Hoeft}, {Horneffer}, {Iacobelli}, {Intema}, {Juette},
  {Karastergiou}, {Kondratiev}, {Kramer}, {Kuniyoshi}, {Kuper}, {van Leeuwen},
  {Loose}, {Maat}, {Mann}, {Markoff}, {McFadden}, {McKay-Bukowski}, {McKean},
  {Mevius}, {Mulcahy}, {Munk}, {Norden}, {Orru}, {Paas}, {Pandey-Pommier},
  {Pandey}, {Pietka}, {Pizzo}, {Polatidis}, {Reich}, {R{\"o}ttgering},
  {Scaife}, {Schwarz}, {Serylak}, {Sluman}, {Smirnov}, {Stappers}, {Steinmetz},
  {Stewart}, {Swinbank}, {Tagger}, {Tang}, {Tasse}, {Toribio}, {Vermeulen},
  {Vocks}, {Vogt}, {van Weeren}, {Wijers}, {Wijnholds}, {Wise}, {Wucknitz},
  {Yatawatta}, {Zarka}, \& {Zensus}}]{buitink2016}
{Buitink} S. {et~al.}, 2016, \nat{}, 531, 70

\bibitem[{{Corstanje} {et~al}\mbox{.}(2016){Corstanje}, {Buitink}, {Enriquez},
  {Falcke}, {H{\"o}randel}, {Krause}, {Nelles}, {Rachen}, {Schellart},
  {Scholten}, {ter Veen}, {Thoudam}, \& {Trinh}}]{Corstanje2016}
{Corstanje} A. {et~al.}, 2016, Astronomy \& Astrophysics, 590, A41

\bibitem[{Hare {et~al}\mbox{.}(2019)Hare {et~al.}}]{Hare2018}
Hare B., {et~al.}, 2019, Nature, 568, 360

\bibitem[{{Huege} {et~al}\mbox{.}(2015){Huege}, {Bray}, {Buitink}, {Dallier},
  {Ekers}, {Falcke}, {James}, {Martin}, {Revenu}, {Scholten}, \&
  {Schroeder}}]{2015aska.confE.148H}
{Huege} T. {et~al.}, 2015, Advancing Astrophysics with the Square Kilometre
  Array (AASKA14), 148

\bibitem[{James(2009)}]{cwjthesis}
James C.~W., 2009, PhD thesis, School of Chemistry and Physics : Physics and
  Mathematical Physics, University of Adelaide

\bibitem[{{James} {et~al}\mbox{.}(2010){James}, {Ekers},
  {{\'A}lvarez-Mu{\~n}iz}, {Bray}, {McFadden}, {Phillips}, {Protheroe}, \&
  {Roberts}}]{2010PhRvD..81d2003J}
{James} C.~W., {Ekers} R.~D., {{\'A}lvarez-Mu{\~n}iz} J., {Bray} J.~D.,
  {McFadden} R.~A., {Phillips} C.~J., {Protheroe} R.~J., {Roberts} P., 2010,
  \prd, 81, 042003

\bibitem[{{Kocz} {et~al}\mbox{.}(2015){Kocz}, {Greenhill}, {Barsdell}, {Price},
  {Bernardi}, {Bourke}, {Clark}, {Craig}, {Dexter}, {Dowell}, {Eftekhari},
  {Ellingson}, {Hallinan}, {Hartman}, {Jameson}, {MacMahon}, {Taylor},
  {Schinzel}, \& {Werthimer}}]{2015Kocz}
{Kocz} J. {et~al.}, 2015, Journal of Astronomical Instrumentation, 4, 1550003

\bibitem[{Krause(2013)}]{Krause2013}
Krause M., 2013, {Master Thesis: Calibration of the LOFAR Antennas, Radboud
  University, 2013}

\bibitem[{{Monroe} {et~al}\mbox{.}(2019){Monroe} {et~al.}}]{Monroe2018}
{Monroe} R., {et~al.}, 2019, Submitted to Nuclear Instrument and Methods in
  Physics Research A, currently published as: Monroe, Ryan McKay (2018)
  Gigahertz Bandwidth and Nanosecond Timescales: New Frontiers in Radio
  Astronomy Through Peak Performance Signal Processing. Dissertation (Ph.D.),
  California Institute of Technology. doi:10.7907/25DP-J474

\bibitem[{Mulrey {et~al}\mbox{.}(2019)Mulrey {et~al.}}]{Mulrey2018}
Mulrey K., {et~al.}, 2019, Astropart. Phys., 111, 1

\bibitem[{{Nehls} {et~al}\mbox{.}(2008){Nehls}, {Hakenjos}, {Arts},
  {Bl{\"u}mer}, {Bozdog}, {van Cappellen}, {Falcke}, {Haungs}, {Horneffer},
  {Huege}, {Isar}, \& {Kr{\"o}mer}}]{LOPES2008}
{Nehls} S. {et~al.}, 2008, Nuclear Instruments and Methods in Physics Research
  A, 589, 350

\bibitem[{{Nelles} {et~al}\mbox{.}(2015{\natexlab{a}}){Nelles}, {H{\"o}randel},
  {Karskens}, {Krause}, {Buitink}, {Corstanje}, {Enriquez}, {Erdmann},
  {Falcke}, {Haungs}, {Hiller}, {Huege}, {Krause}, {Link}, {Norden}, {Rachen},
  {Rossetto}, {Schellart}, {Scholten}, {Schr{\"o}der}, {ter Veen}, {Thoudam},
  {Trinh}, {Weidenhaupt}, {Wijnholds}, {Anderson}, {B{\"a}hren}, {Bell},
  {Bentum}, {Best}, {Bonafede}, {Bregman}, {Brouw}, {Br{\"u}ggen}, {Butcher},
  {Carbone}, {Ciardi}, {de Gasperin}, {Duscha}, {Eisl{\"o}ffel}, {Fallows},
  {Frieswijk}, {Garrett}, {van Haarlem}, {Heald}, {Hoeft}, {Horneffer},
  {Iacobelli}, {Juette}, {Karastergiou}, {Kohler}, {Kondratiev}, {Kuniyoshi},
  {Kuper}, {van Leeuwen}, {Maat}, {McFadden}, {McKay-Bukowski}, {Orru}, {Paas},
  {Pandey-Pommier}, {Pandey}, {Pizzo}, {Polatidis}, {Reich}, {R{\"o}ttgering},
  {Schwarz}, {Serylak}, {Sluman}, {Smirnov}, {Tasse}, {Toribio}, {Vermeulen},
  {van Weeren}, {Wijers}, {Wucknitz}, \& {Zarka}}]{Nelles2015}
{Nelles} A. {et~al.}, 2015{\natexlab{a}}, Journal of Instrumentation, 10,
  P11005

\bibitem[{{Nelles} {et~al}\mbox{.}(2015{\natexlab{b}}){Nelles}, {Schellart},
  {Buitink}, {Corstanje}, {de Vries}, {Enriquez}, {Falcke}, {Frieswijk},
  {H{\"o}randel}, {Scholten}, {ter Veen}, {Thoudam}, {van den Akker},
  {Anderson}, {Asgekar}, {Bell}, {Bentum}, {Bernardi}, {Best}, {Bregman},
  {Breitling}, {Broderick}, {Brouw}, {Br{\"u}ggen}, {Butcher}, {Ciardi},
  {Deller}, {Duscha}, {Eisl{\"o}ffel}, {Fallows}, {Garrett}, {Gunst},
  {Hassall}, {Heald}, {Horneffer}, {Iacobelli}, {Juette}, {Karastergiou},
  {Kondratiev}, {Kramer}, {Kuniyoshi}, {Kuper}, {Maat}, {Mann}, {Mevius},
  {Norden}, {Paas}, {Pandey-Pommier}, {Pietka}, {Pizzo}, {Polatidis}, {Reich},
  {R{\"o}ttgering}, {Scaife}, {Schwarz}, {Smirnov}, {Stappers}, {Steinmetz},
  {Stewart}, {Tagger}, {Tang}, {Tasse}, {Vermeulen}, {Vocks}, {van Weeren},
  {Wijnholds}, {Wucknitz}, {Yatawatta}, \& {Zarka}}]{Nelles2014}
{Nelles} A. {et~al.}, 2015{\natexlab{b}}, Astroparticle Physics, 65, 11

\bibitem[{{Schellart} {et~al}\mbox{.}(2013){Schellart}, {Nelles}, {Buitink},
  {Corstanje}, {Enriquez}, {Falcke}, {Frieswijk}, {H{\"o}randel}, {Horneffer},
  {James}, {Krause}, {Mevius}, {Scholten}, {ter Veen}, {Thoudam}, {van den
  Akker}, {Alexov}, {Anderson}, {Avruch}, {B{\"a}hren}, {Beck}, {Bell},
  {Bennema}, {Bentum}, {Bernardi}, {Best}, {Bregman}, {Breitling}, {Brentjens},
  {Broderick}, {Br{\"u}ggen}, {Ciardi}, {Coolen}, {de Gasperin}, {de Geus}, {de
  Jong}, {de Vos}, {Duscha}, {Eisl{\"o}ffel}, {Fallows}, {Ferrari}, {Garrett},
  {Grie{\ss}meier}, {Grit}, {Hamaker}, {Hassall}, {Heald}, {Hessels}, {Hoeft},
  {Holties}, {Iacobelli}, {Juette}, {Karastergiou}, {Klijn}, {Kohler},
  {Kondratiev}, {Kramer}, {Kuniyoshi}, {Kuper}, {Maat}, {Macario}, {Mann},
  {Markoff}, {McKay-Bukowski}, {McKean}, {Miller-Jones}, {Mol}, {Mulcahy},
  {Munk}, {Nijboer}, {Norden}, {Orru}, {Overeem}, {Paas}, {Pandey-Pommier},
  {Pizzo}, {Polatidis}, {Renting}, {Romein}, {R{\"o}ttgering}, {Schoenmakers},
  {Schwarz}, {Sluman}, {Smirnov}, {Sobey}, {Stappers}, {Steinmetz}, {Swinbank},
  {Tang}, {Tasse}, {Toribio}, {van Leeuwen}, {van Nieuwpoort}, {van Weeren},
  {Vermaas}, {Vermeulen}, {Vocks}, {Vogt}, {Wijers}, {Wijnholds}, {Wise},
  {Wucknitz}, {Yatawatta}, {Zarka}, \& {Zensus}}]{schellart2013}
{Schellart} P. {et~al.}, 2013, Astronomy \& Astrophysics, 560, A98

\bibitem[{{The Pierre Auger Collaboration}(2016)}]{Auger2016}
{The Pierre Auger Collaboration}, 2016, Journal of Instrumentation, 11, P01018

\bibitem[{{Thoudam} {et~al}\mbox{.}(2014){Thoudam}, {Buitink}, {Corstanje},
  {Enriquez}, {Falcke}, {Frieswijk}, {H{\"o}randel}, {Horneffer}, {Krause},
  {Nelles}, {Schellart}, {Scholten}, {ter Veen}, \& {van den
  Akker}}]{Thoudam2014}
{Thoudam} S. {et~al.}, 2014, Nuclear Instruments and Methods in Physics
  Research A, 767, 339

\bibitem[{{Wilson} {et~al}\mbox{.}(2011){Wilson}, {Ferris}, {Axtens}, {Brown},
  {Davis}, {Hampson}, {Leach}, {Roberts}, {Saunders}, {Koribalski}, {Caswell},
  {Lenc}, {Stevens}, {Voronkov}, {Wieringa}, {Brooks}, {Edwards}, {Ekers},
  {Emonts}, {Hindson}, {Johnston}, {Maddison}, {Mahony}, {Malu}, {Massardi},
  {Mao}, {McConnell}, {Norris}, {Schnitzeler}, {Subrahmanyan}, {Urquhart},
  {Thompson}, \& {Wark}}]{2011MNRAS.416..832W}
{Wilson} W.~E. {et~al.}, 2011, \mnras, 416, 832

\end{thebibliography}

\end{document}